\documentclass[twocolumn]{aastex701}

\usepackage{lipsum}      
\usepackage{graphicx}
\usepackage{amssymb}
\usepackage{amsmath}
\usepackage{epstopdf,hyperref}
\usepackage{xcolor}
\usepackage{float}

\newcommand{\beq}{\begin{equation}}
\newcommand{\eeq}{\end{equation}}

\begin{document}

\title{Inferring the Dark from the Observable: Estimating Halo Masses Using Galaxy Properties}

\author[orcid=0000-0003-4780-5554,sname='Chen']{Alice Chen}
\email[show]{ay7chen@uwaterloo.ca}
\affiliation{Perimeter Institute for Theoretical Physics, 31 Caroline Street North, Waterloo, Ontario N2L 2Y5, Canada}
\affiliation{Department of Physics and Astronomy, University of Waterloo, 200 University Avenue West, Waterloo, Ontario N2L 3G1, Canada}
\affiliation{Waterloo Centre for Astrophysics, University of Waterloo, 200 University Avenue West, Waterloo, Ontario N2L 3G1, Canada}

\author[orcid=0000-0001-6336-642X,sname='Ahad']{Syeda Lammim Ahad}
\affiliation{Waterloo Centre for Astrophysics, University of Waterloo, 200 University Avenue West, Waterloo, Ontario N2L 3G1, Canada}
\affiliation{Department of Physics and Astronomy, University of Waterloo, 200 University Avenue West, Waterloo, Ontario N2L 3G1, Canada}
\affiliation{Center for Astronomy, Space Science and Astrophysics, Independent University, Bangladesh, Dhaka 1229, Bangladesh}
\email{slahad@uwaterloo.ca}

\begin{abstract}
The formation and evolution of galaxies are interconnected with that of their host halos.  Given this closely related evolutionary history, it should follow that galaxy properties are correlated with their host halo mass.  Previous works have shown that star formation rates of galaxies in cluster halos differ systematically from those in field halos, suggesting that host halo mass plays a significant role in the history of galaxy evolution.  Here, we examine the correlations between observable galaxy properties - such as stellar mass, star formation rate, half-stellar radius, $r-$band magnitude, and color - and the underlying host halo mass of galaxies.  Central galaxy observables are used to probe host halo mass, while satellite galaxy properties are used to estimate subhalo mass.  We perform this analysis using random forest, ordinary least squares (OLS), and symbolic regression.  Random forest regression can assess the relative significance of different observable galaxy properties, while OLS and symbolic regression can quantify their relationship with the host halo mass.  Our results show that gas mass is generally the most significant property in central galaxies of all halo mass ranges, followed by stellar mass and observed $r-$band magnitude.  Adding parameters such as the colour and magnitude of galaxies improves host halo mass estimations compared to standard stellar-to-halo mass relations.  These results show that halo mass likely has a multi-dimensional dependence on galaxy properties and open the avenue of finding a simple way to constrain halo mass based on what is directly observable in galaxy surveys.  Finding the observables that have the strongest correlation with halo mass can also motivate future surveys on which areas to concentrate in detail.
\end{abstract}

\keywords{\uat{Galaxy dark matter halos}{1880} --- \uat{Galaxy masses}{607} --- \uat{Galaxies}{573} --- \uat{Cosmology}{343} --- \uat{Magnetohydrodynamical simulations}{1966} --- \uat{Hydrodynamical simulations}{767}}

\section{Introduction}
\label{sec:intro}
The cosmology of the universe is currently best parameterized by the $\Lambda$CDM model, where dark matter is cold and collisionless.  This favors the hierarchical model of structure formation, where galaxies first form within dark matter potential wells \citep{white1978} and undergo mergers to form the structures we see in the present day \citep{Cooray:2002dia}.  In the early Universe, primordial overdensities caused by quantum fluctuations grow via gravitational instability, which deepens the local gravitational potential wells.  Baryons are then pulled into these potential wells, and eventually fall into the centers of these wells \citep{white1978}.  When the dark matter overdensity reaches the critical value of $\sim 180 \Delta_{crit}$, where $\Delta_{crit} = \Omega_m \Bar{\rho}$ and $\Bar{\rho}$ is the average density of the universe, the overdense dark matter region collapses to form virialized structures known as halos \citep{Cooray:2002dia, Asgari:2023mej}.  The baryonic component in the halo center undergoes a similar collapse process, forming the progenitor of what we observe as the galaxies in the center of their halos in the Local Universe (redshift $\leq 0.05$).  The galaxies of these halos in present day are known as centrals.  Subsequent halo mergers and accretion drive the growth of both the dark matter halo and its central, assembling into large-scale structure we see today \citep{Mo2010Galaxy, Dalal:2008zd}.  The satellite galaxies in halos come from these mergers - they were originally centrals of their own halos before being merged into the larger halo \citep{white1978, Simha:2008hd}.

From this formation scenario, it is clear that galaxies formed and co-evolved alongside their host halos.  This co-evolution implies that observable galaxy properties should retain correlations with its host halo properties, such as mass, spin, and formation timescale.  The property that we will be focusing on here is halo mass, since it is one of the fundamental properties of the halo. 

Currently the most common way to determine the total halo mass is through weak lensing \citep{Mandelbaum:2017jpr, Li:2024ogc}.  However, this is not a precise measurement because of factors such as galaxy shape noise, projection effects, selection criterion, and source redshift uncertainty - all of which introduce systematic errors in the final measurement \citep{Nde:2025dzf}.  This lack of precision increases for lower mass objects, particularly subhalos, which makes determining subhalo mass by lensing often unfeasible.  This motivates the need for alternative methods to probe total halo mass.

Here, we aim to explore the galaxy-halo connection by quantifying the correlation between observable galactic properties and the galaxies' host halo mass, as well as isolating the importance of each galactic property in relation to the host mass (see Table \ref{tab:importances}).  Evidence for the correlation between observables and host halo mass is seen in the results of \citet{Shankar_2006, Bilicki:2021hgn, 2023MNRAS.518.1002R, 2023OJAp....6E..37B, ahad2024, 2025arXiv251019926D}.  Focusing on galaxy properties is important because the goal is to be able to quantify host halo masses in observations in real space with measurable data.  

We focus on directly observable properties, such as stellar mass, star formation rate, half-stellar radius, $r-$band magnitude, and $g-r$ color, since from theory, a halo-to-stellar mass relation exists \citep{1976ApJ...203..297S, Cooray:2002dia} and galaxy color has been found to have a correlation with host halo mass \citep{2021NatAs...5.1069C}.  The gas content of a halo is another key property.  However, the gas mass is not easy to obtain, though the thermal Sunyaev-Zel'dovich (SZ) effect \citep{Nagai:2005wx, 10.1093/mnras/stac2505} and data from radio observations can constrain gas mass in galaxies.  We also include the formation redshift, which is the redshift at which the halo first assembled half of its current total mass.  This is important for assembly bias and galaxy-halo formation history; the formation history should have some effect on the host mass as well \citep{Dalal:2008zd}.  Values of dark matter concentration \footnote{Navarro Frenk White (NFW) concentration, $c=\frac{r_{vir}}{r_{scale}}$ \citep{1996ApJ...462..563N}} and assembly bias are key quantities often cited in more complex scaling relations related to the dark matter distribution of halos \citep{Cooray:2002dia, Asgari:2023mej}; however, they are often unobservable or have ambiguous definitions, so we use different mass components, half-mass radius, and formation redshift as proxies here.

Quantifying the mass of the host halo is useful because we can learn how much dark matter is expected to be around a galaxy and give insight to that galaxy's evolutionary history.  This also helps constrain the total mass content in the universe for cosmology.  Late time tracers, such as weak lensing, and early time tracers, such as the CMB, currently do not match each other, leading to a $\sigma_8$ tension.  A better estimation of host halo masses can reduce late time systematics of $\sigma_8$ measurements \citep{Abdalla:2022yfr, Karim:2024luk}.  

As simulations and observations have improved significantly over the past few decades, it has become necessary to account for multi-dimensional scaling relations to estimate the host halo mass compared to the simpler luminosity–mass relation often used \citep{1976ApJ...203..297S, 2025arXiv251019926D, huang2026tighterdarkmatterconstraints}.  Due to the amount of unknowns and low signal-to-noise ratio of observational data, we focused on simulation data for this multi-dimensional analysis.  We use TNG-100 simulation \citep{Nelson:2018uso}, which is publicly available and has high enough resolution for baryonic physics in galaxies as well as available parameters for star formation, different component masses (dark matter, gas, stellar, etc), and half-stellar radius.  These quantities are known values in simulations and are not subject to systematic errors or foreground contaminations like the ones found in observations, making them more useful for finding clear correlations between data compared to observational data, where even ``observable" quantities like stellar mass and star formation rate are estimated based on luminosity.

This paper is structured as follows - section \ref{sec:data_method} specifies the data and techniques we use for quantitative analysis, section \ref{sec:results} outlines our findings and estimation of the host halo mass, section \ref{sec:discussion} discusses potential underlying physics behind our results and further error analysis, and section \ref{sec:conclusion} summarizes this paper and suggests future work.  A flat $\Lambda$CDM cosmology is assumed for any relevant calculations in this work, with Planck 2015 parameters \citep{Planck:2015fie}.

\section{Data and Methods}
\label{sec:data_method}

\subsection{Simulation Data}\label{subsec:data}

The simulation data used for this analysis are from TNG100 in the Illustris-TNG simulations suite \citep{Nelson:2018uso, Vogelsberger:2014dza}).  This simulation has a box side of 75Mpc/h, implements Planck 2015 cosmology \citep{Planck:2015fie}, and is built upon the cosmological simulation code AREPO \citep{2010MNRAS.401..791S}.  AREPO finds the magneto-hydrodynamical equations in a quasi-Lagrangian fashion -  to simulate the baryonic physics of the galaxies in the simulations - and calculates gravitational forces with a Tree-Particle-Mesh (Tree-PM) scheme \citep{Pillepich:2017jle}.

\subsubsection{Sample Selection of Halos and Galaxies} \label{sec:method_data_subselect}

We divide the host halo sample into two categories: field halos that are $10^{12}\text{M}_{\odot} \leq M_{field} \leq 10^{13}$ and group + cluster halos that have $M_{group} \geq 10^{13} \text{M}_{\odot}$, where $M_{field}$ and $M_{group}$ are the total FOF halo masses.  We use FOF mass instead of $M_{virial}$ or $M_{200c}$ because we want to estimate the total mass of the host halos, which may not follow NFW profiles \citep{Salucci_2019} or be spherical, and the FOF mass is the closest proxy.  We also split the halos into different categories because the halo environment likely plays a key role in its mass accretion history and evolution, so correlations between observable galaxy properties could differ between field and group/cluster sized halos \citep{Kauffmann:2004sj, Dalal:2008zd, Skibba_2015}.  There are 1988 halos in our field sample and 215 in group+cluster sample.

Although the latter category could be further split into groups and clusters individually, only 17 cluster halos ($M_{halo} \geq 10^{14} \text{M}_{\odot}$) are present in the data (with 198 group sized halos), making their results statistically insignificant.  Therefore, the clusters were combined with the group halos in the analysis below.

The primary objective here is to quantify how the underlying dark matter mass of the halo depends on the observable properties of its galaxies.  We split the galaxies based on central and satellite classification, with the central properties being used to estimate host halo mass and the satellites being used to estimate subhalo mass.

To ensure that the galaxies used in our analysis are physically realistic, we first removed all subhalos flagged by the TNG simulation as numerical artefacts.  We then enforced a selection criterion where the stellar mass contained within a 30kpc aperture is at most $20\%$ of the total subhalo mass, close to the minimum bound of the luminosity-to-mass relation \citep{2010ApJ...717..379B, Wechsler:2018pic}, so that the subhalos in the analysis have a nonzero distribution of gas, dark matter, and stars.  We use a 30kpc aperture because the 3D stellar mass within this aperture from simulations is comparable to the measured stellar mass within the Petrosian radius of observed galaxies \citep{2015MNRAS.446..521S, Pillepich:2017jle}.  The fraction of a halo's mass contained in stars is typically significantly less than its gas + dark matter component \citep{2010ApJ...717..379B, 2018AstL...44....8K}.  If the total mass of a galaxy comes solely from its stellar mass, then it is devoid of dark matter, and thus has a non-standard mass-to-light ratio.  This would not be representative of typical galaxies in observations, and given both numerical and theoretical limitations, it is unclear whether this non-standard mass-to-light ratio is physical or a numerical error, so these galaxies were excluded from our analysis.  Furthermore, we also imposed a stellar mass cutoff of $\geq 10^9 M_{\odot}$ for the galaxies in our selection.  This ensured each galaxy was well resolved, with $\geq 1000$ star particles, given the simulation resolution limit.  The resulting subhalos have the stellar mass-total subhalo mass relation shown in Figure \ref{fig:aperture_mass} below. 

\begin{figure*}
    \centering
    \includegraphics[width=0.8\linewidth]{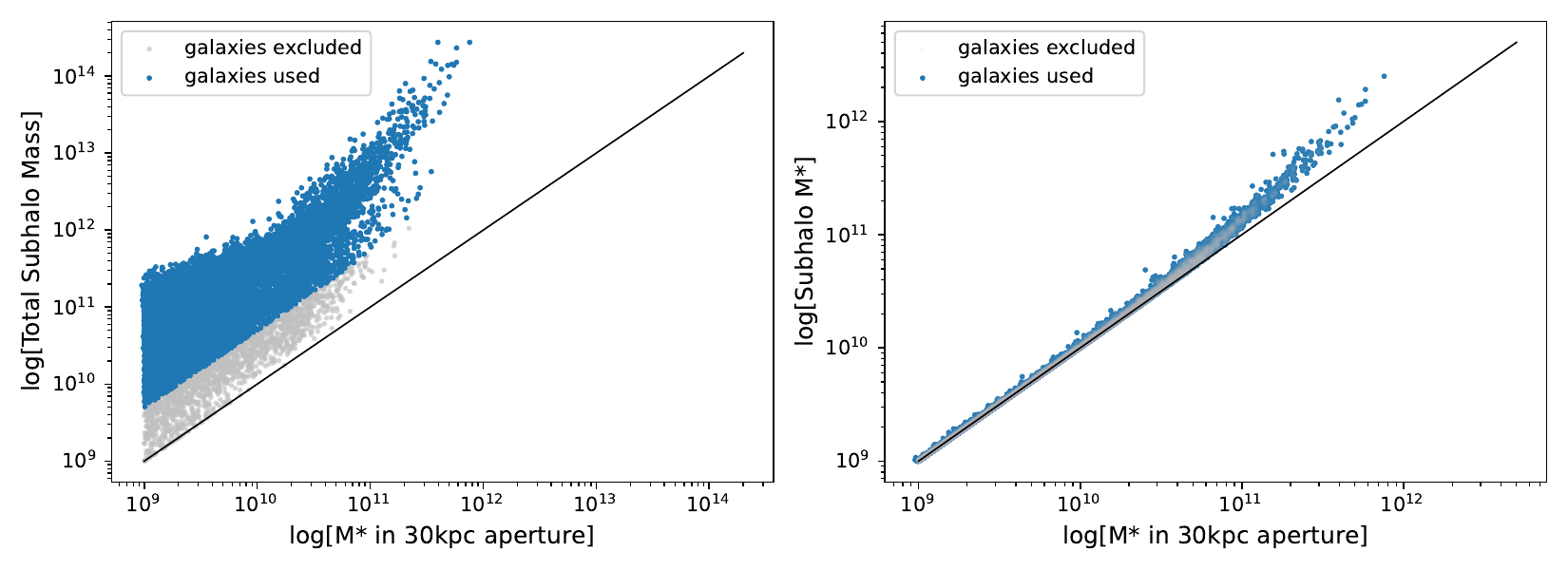}
    \caption{\footnotesize Aperture mass vs total subhalo mass and stellar mass.  The subhalos in this sample already have a constraint of stellar mass$\geq 10^9 \text{M}_{\odot}$.  There are no points that significantly deviate from the trend of approximate linear relation between mass and aperture mass, so the subhalos in the simulation are already distinct from each other and we don't need to apply further cuts here.  On the right subplot, we can see that the stellar mass in a 30kpc aperture $\leq$ total galaxy mass, which is physically consistent, since 
    $10^9 + \text{M}_{\odot}$ galaxies typically have radii $\geq$ 30kpc.}
    \label{fig:aperture_mass}
\end{figure*}

\subsubsection{Galaxy Properties}
From the selected galaxy sample, we examine galaxy properties within each halo type (centrals, satellites, field, groups, and clusters).  We use these galaxy properties as the training parameters, and split them into the following set and subsets to be trained on:
\begin{itemize}
    \item All properties: stellar mass (units $\text{M}_{\odot}$), gas mass (units $\text{M}_{\odot}$), half-stellar radius (units kpc/h), specific star formation rate (sSFR) (absolute SFR ($M_{\odot}$/year)/total $M_{\ast}$), formation redshift, $r-$band magnitude, $g-r$ color
    \item Observables + gas mass: stellar mass, gas mass, half-stellar radius, sSFR, formation redshift
    \item Observable properties only: stellar mass, half-stellar radius, sSFR, $r-$band magnitude, $g-r$ color.
    \label{list:properties}
\end{itemize}

Stellar mass and gas mass here are the total masses of the stars and gas within the FOF group (host halo) for central galaxies, and subhalo (``sub-group") for satellites.  The half-stellar radius (also known as the stellar half-mass radius) is the radius containing half of the total stellar mass in the galaxy (central when estimating host halo mass, satellite when estimating subhalo mass).  The sSFR measurement here is the specific star formation rate, calculated by 
\beq
sSFR = \frac{\text{absolute star formation rate}}{M_{\ast}},
\eeq 
where absolute star formation rate is the SFR in the simulations in units of $M_{\ast}/$year, and $M_{\ast}$ is the total stellar mass of the galaxy.  The color and magnitudes from the simulations are matched to galaxies of comparable stellar masses in the SDSS catalog \citep{Nelson:2017cxy, Pillepich:2017jle}.

\subsubsection{Evaluation of Galaxy Properties}
\begin{figure*}
    \centering    
    \includegraphics[trim=50 0 50 0,clip, width=\linewidth]{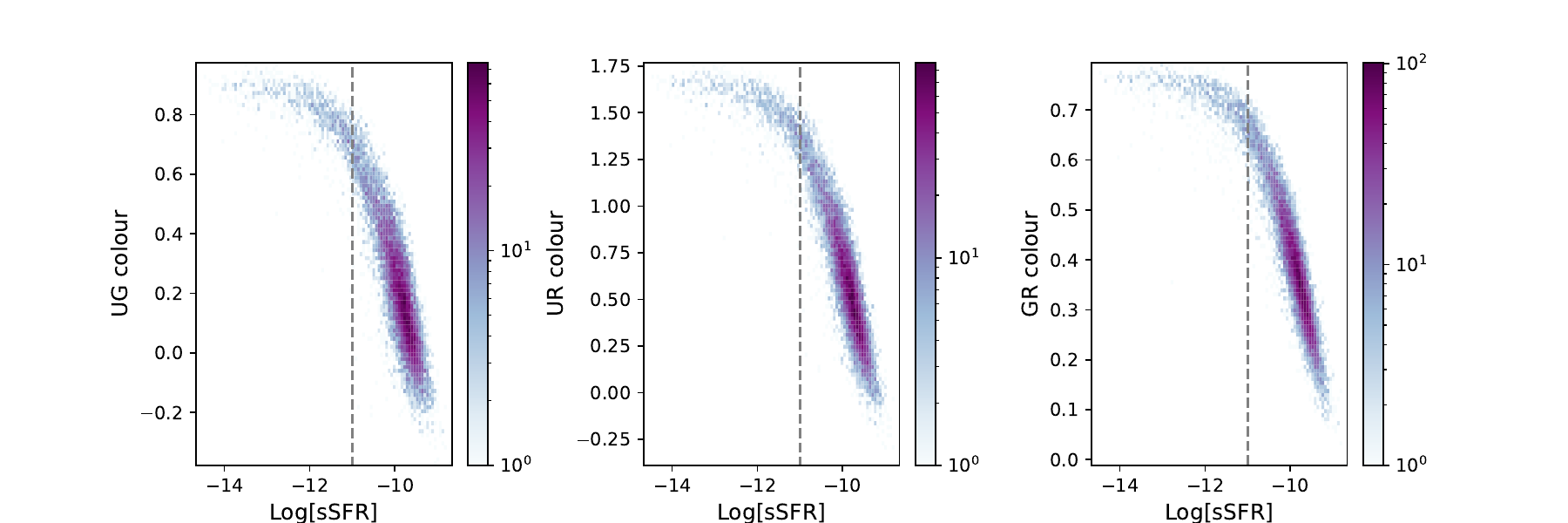}
    \caption{\footnotesize Plot of log specific star formation rate vs. color.  As expected from theory, more star-forming galaxies are brighter in bluer bands.}
    \label{fig:colour_sfr}
\end{figure*}

To further ensure that the data behaves physically, we look at how the colors and magnitudes of galaxies are correlated with star formation rates, shown in Figure \ref{fig:colour_sfr}.  This is for our entire galaxy sample, including centrals and satellites of all halo masses.  While the magnitudes in the Illustris-TNG simulations seem to be 0.5-1.0 dex brighter compared to galaxies at similar stellar masses in observational catalogs, \citep{Alsing:2024tlr, Thorp:2025bpl}, they are important observables and we want to see how their behavior follows sSFR.  Across all panels in Fig. \ref{fig:colour_sfr}, we can see that higher star formation rates correspond to bluer colors (smaller values on the y-axis in Figure \ref{fig:colour_sfr}), which is consistent with stellar astrophysics, since younger, newly formed stars tend to be bluer in color.  From this, we decided to keep the $r-$band magnitude and $g-r$ color properties in our analysis below. 

\subsection{Machine Learning Techniques}
With this galaxy and halo sample, we develop a systematic method to quantify the host halo mass of galaxies based on their observable properties by using machine learning algorithms such as random forest and symbolic regression.  We start with random forest regression to assess the feature importance of galaxy properties, then quantify the scaling relation between properties and host halo mass using random forest regression, ordinary least squares (OLS), and symbolic regression.  Throughout, we split the galaxies into a testing and training sample; the galaxies are randomized, 75$\%$ are selected as the training sample, and the remaining 25$\%$ are the testing sample.  Furthermore, the training parameters were normalized by dividing by the largest value of each parameter in the sample (e.g. $M_{\ast}/M_{\ast, max}$) to prevent skewing results due to differences in parameter scaling.  This train/test split and normalization was done for all ML methods below.

\subsubsection{Random Forest Regression}
We start by employing a random forest regression algorithm to assess the relative importance of different observable galaxy properties in predicting the host halo mass.  Random forests assess the nonlinear correlations between multiple features and a single outcome \citep{breiman2001random} - in this case, the features are the galaxy properties and the outcome is the total host halo mass.  They construct multiple decision trees and aggregate their predictions, they are useful for providing a natural measure of feature significance.  This is one of the two steps we take in determining which quantities to focus on in predicting the total dark matter mass of the host halo; the other step being symbolic regression (see Sec.~\ref{subsec:symbolic_regression} below).  We also use a random forest regressor to predict host halo mass using different parameter sets to compare the purely ML results to symbolic formula predictions later on (see Tables \ref{tab:importances} and \ref{tab:eqns}, respectively).

\subsubsection{Ordinary Least Squares}
\label{subsec:ols}
Random forest decision trees can provide insight into the galaxy parameters that have the most correlation with host halo mass, but do not give a mathematical scaling relation.  To estimate this scaling to first order, we start by fitting a multivariate linear model using ordinary least squares (OLS).  OLS finds linear relations between each galaxy parameter and the host halo mass by estimating the coefficients in 
\beq 
M_{halo} = a_1x_1 + a_2x_2 + ... + a_nx_n,
\eeq 
where $n$ is the number of galaxy properties in our training set and $x_1, ..., x_n$ are the properties themselves.  The parameters are fitted by minimizing the sum of the squared residuals between the predicted and true halo masses \citep{2022arXiv221115347Z}.
While many of the properties may have a nonlinear relation to the host halo mass instead (see Sec. \ref{subsec:symbolic_regression}), OLS can give a useful preliminary model that is computationally fast and easy to interpret.  Comparing its performance with more flexible nonlinear methods - such as symbolic regression - allows us to assess how well the halo mass can be described by simple linear scaling relations.

\subsubsection{Symbolic Regression}
\label{subsec:symbolic_regression}
Next, we run a symbolic regression algorithm on the observable properties using PySR \citep{Cranmer:2020wew, 2023arXiv230501582C} to see if the same properties appear in order of frequency in the resulting fitting functions.  Symbolic regression is a machine learning algorithm that explores the parameter space of mathematical functions to fit a function to data points inputted by the user.  It is similar to linear regression \citep{Roustaei2024}, but instead of a linear function, it can fit all types of mathematical functions, such as trigonometric functions, exponential/logarithmic functions, and polynomials (see Table \ref{tab:eqns} for reference).

The more frequently a parameter representing an observable property appears in the resulting functions, the more likely it is to be significantly correlated with the underlying host halo mass.  If the PySR equations and random forest feature importances highlight the same variables, then it is a good indicator that the properties corresponding to those variables are strongly correlated with the halo mass, since two independent tests both show the same outcome.  To constrain the functions to be more physical, we restrict the mathematical equations to be a combination of monotonic (e.g. linear, exponential, logarithmic) functions, since dependence of total halo mass likely would not be periodic.  We select the equation with the pareto front that has the highest decrease in error with the lowest increase in complexity, as is standard in symbolic regression.

\section{Halo Mass Prediction Results}
\label{sec:results}
Here, we show how well galaxy properties predict their halo masses.  For central galaxies, the ''halo" refers to the host halo of the system, whereas for satellites, this halo is their individual subhalos.

\subsection{Host Halo Mass for Central Galaxies}
\label{subsec:host_halo_mass_predictions}

In this section, we focus on estimating the total mass of the host halo using their central galaxies only.  We construct distinct training samples from central galaxies in field, group, and cluster-scale environments (see Sec. \ref{sec:method_data_subselect} for environment definitions) and apply the methods described in Sec. \ref{sec:data_method} to estimate their corresponding host halo masses.

\begin{figure}
    \centering    \includegraphics[width=0.85\linewidth]{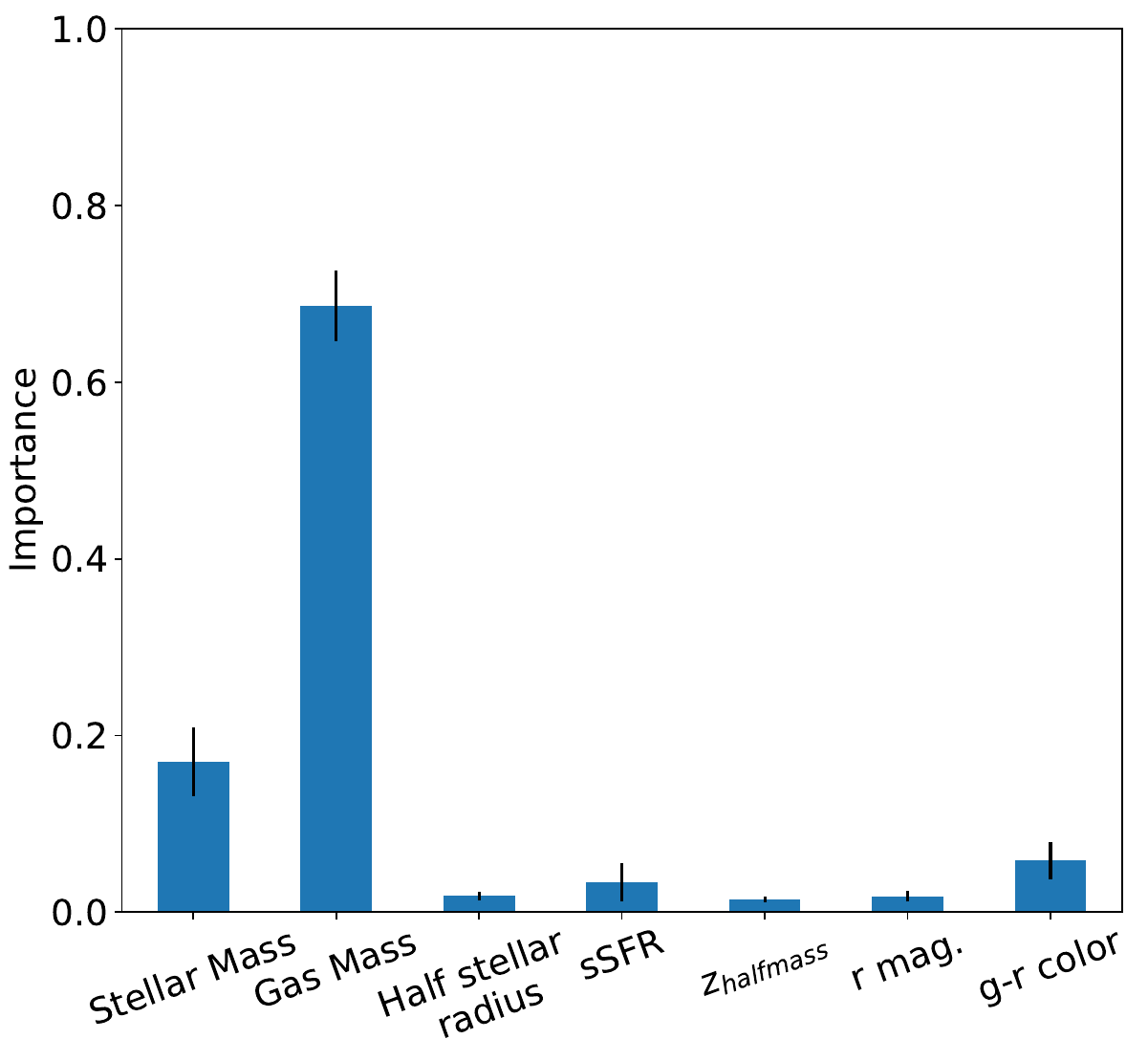} \\
    {\footnotesize a) All features that could be important for field central galaxies.}
        \label{fig:field_rf_all}
        \vspace{2mm}    \includegraphics[width=0.85\linewidth]{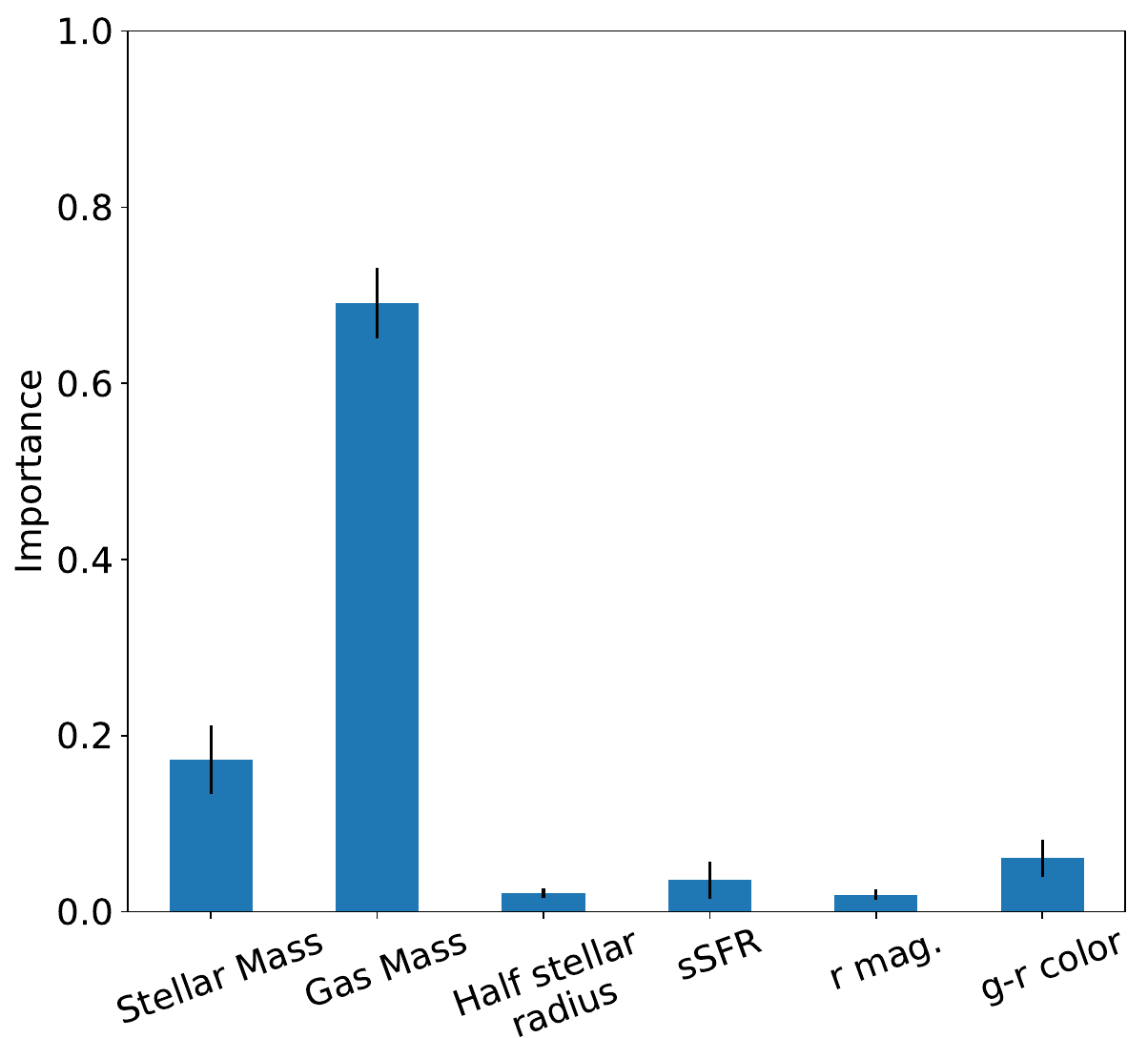} \\
    {\footnotesize b) Features that are found in observations + gas mass.}
        \label{fig:field_rf_sim}
        \vspace{2mm}    \includegraphics[width=0.85\linewidth]{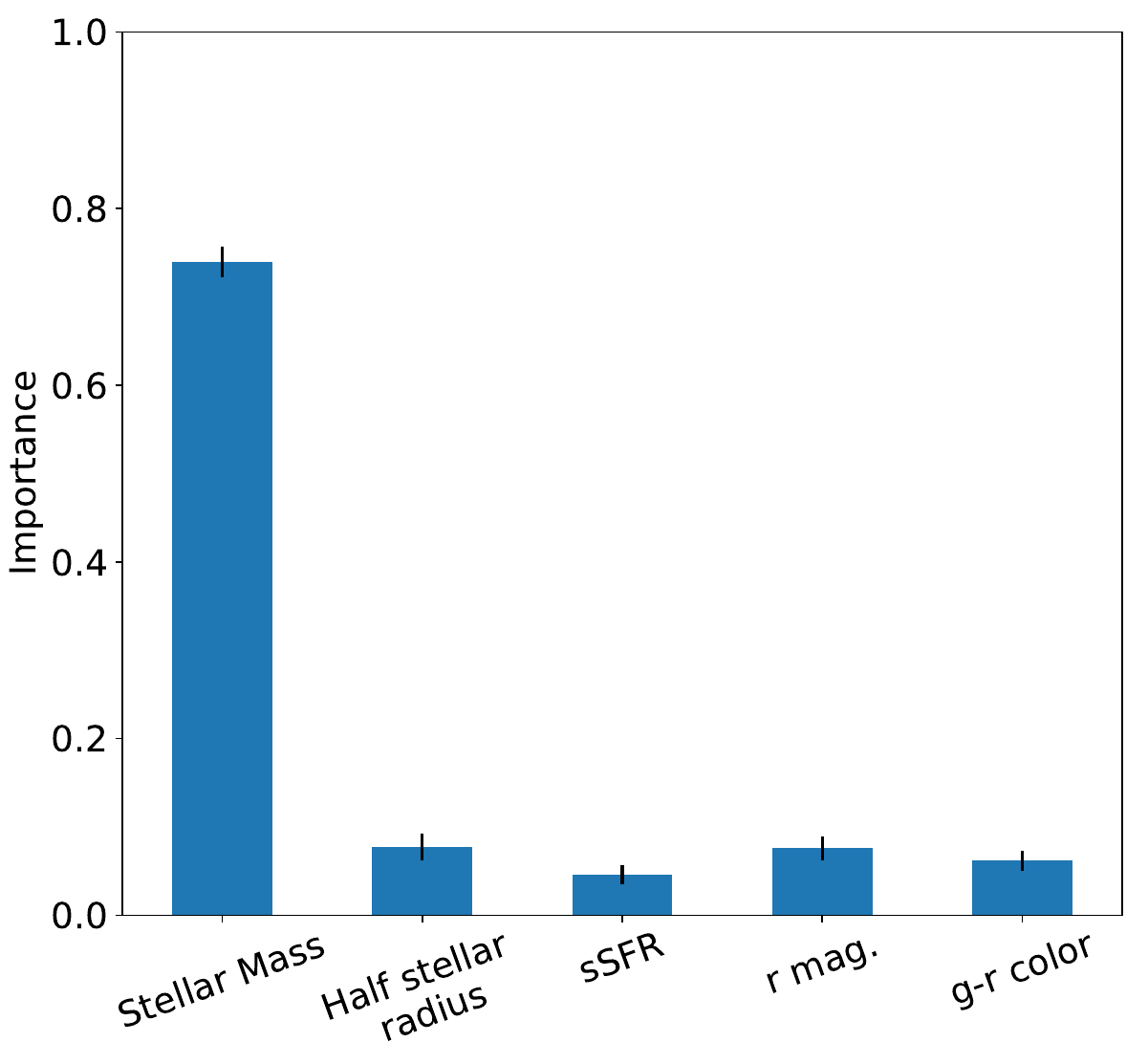} \\
         {\footnotesize c) Features that are in the ``observables only" group.}
         \label{fig:field_rf_obs}
    \caption{\footnotesize Random forest feature importance of different combinations of central galaxy features in field-sized host halos.}
    \label{fig:field_rf}
\end{figure}

\begin{figure}
    \centering
    \includegraphics[width=0.85\linewidth]{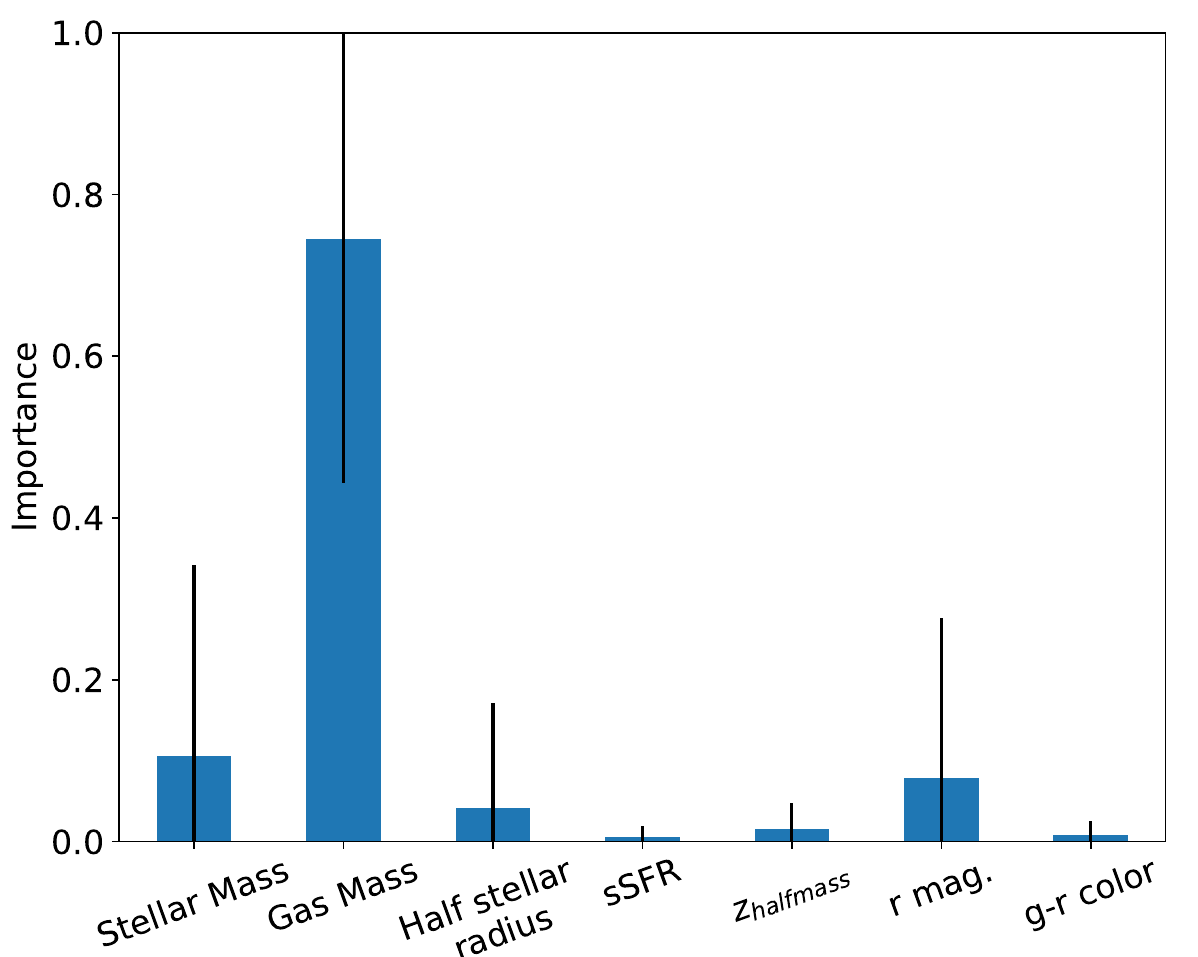}
         \\ {\footnotesize a) All training parameters for central galaxies of group+cluster mass range.}
         \label{fig:gc_rf_all}
         \vspace{2mm}
    \includegraphics[width=0.85\linewidth]{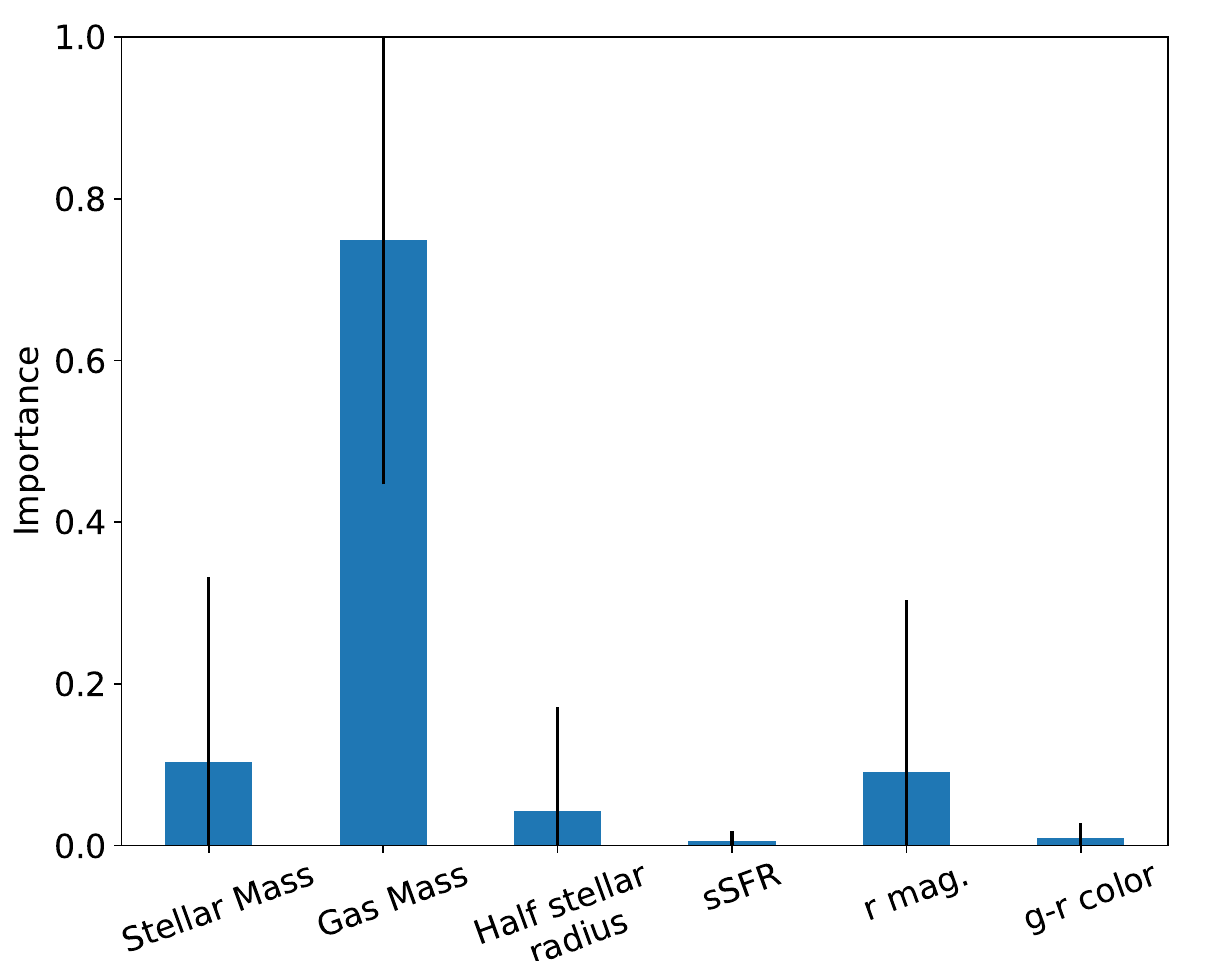}
         \\ {\footnotesize b) Features that are found in simulations.}
         \label{fig:gc_rf_sim}
         \vspace{2mm}
    \includegraphics[width=0.85\linewidth]{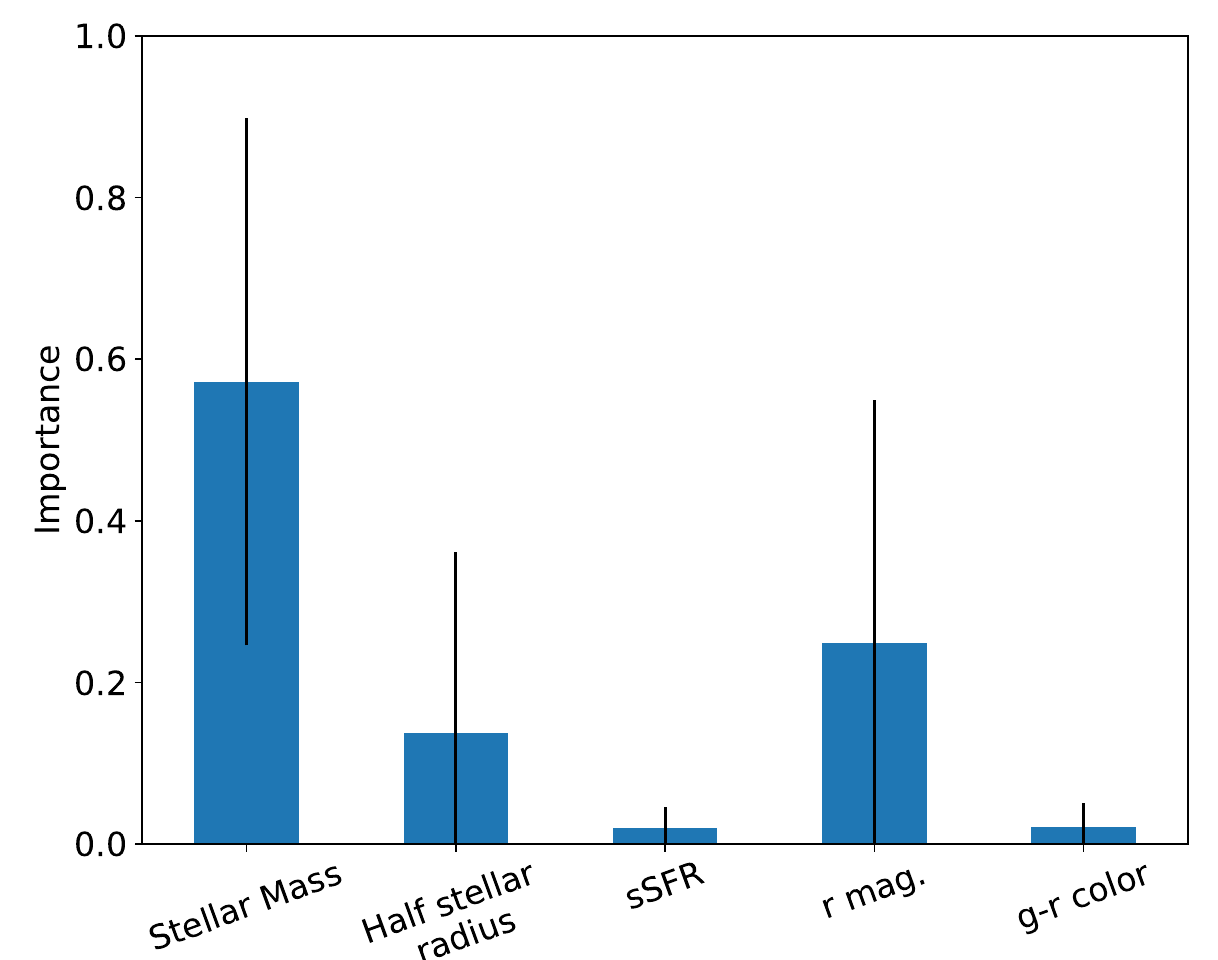}
         \\ {\footnotesize c) Features of the ``observables only" group.}
         \label{fig:gc_rf_obs}
         \vspace{1mm}
        \caption{\footnotesize Random forest feature importance of different combinations of central galaxy features in group and cluster-mass host halos (if the group and cluster samples are combined).}
    \label{fig:gc_rf}
\end{figure}

The results of the random forest regression is obtained from the galaxy parameters in these three categories - all properties, observables + gas mass, observables only - given in the list \ref{list:properties}.  They can be seen in Figure \ref{fig:field_rf} (for the field) and \ref{fig:gc_rf} (for groups/clusters).  In both of these Figures, the a) and b) panels show that gas mass is the most important feature across galaxies in all environments.  This is consistent with the expectation that gas is a significant component of the overall halo mass, since more massive dark matter halos attract more gas into their potential wells.  This would make gas the most massive component of a halo after dark matter - only around $\leq 20\%$ of the accreted gas (baryons) would cool to form stars \citep{Wechsler:2018pic}.  Although gas mass is the most important feature in the group+cluster sample, for the 17 clusters in the TNG-100 simulations, the gas mass is no longer the predominant component, which seems to be unique to this mass bin (see Figure \ref{fig:cluster_rf} in the Appendix).  Given that we only have 17 clusters, it is not statistically significant to draw conclusions about cluster-mass halos.

\begin{figure*}
    \centering
    \includegraphics[width=\linewidth]{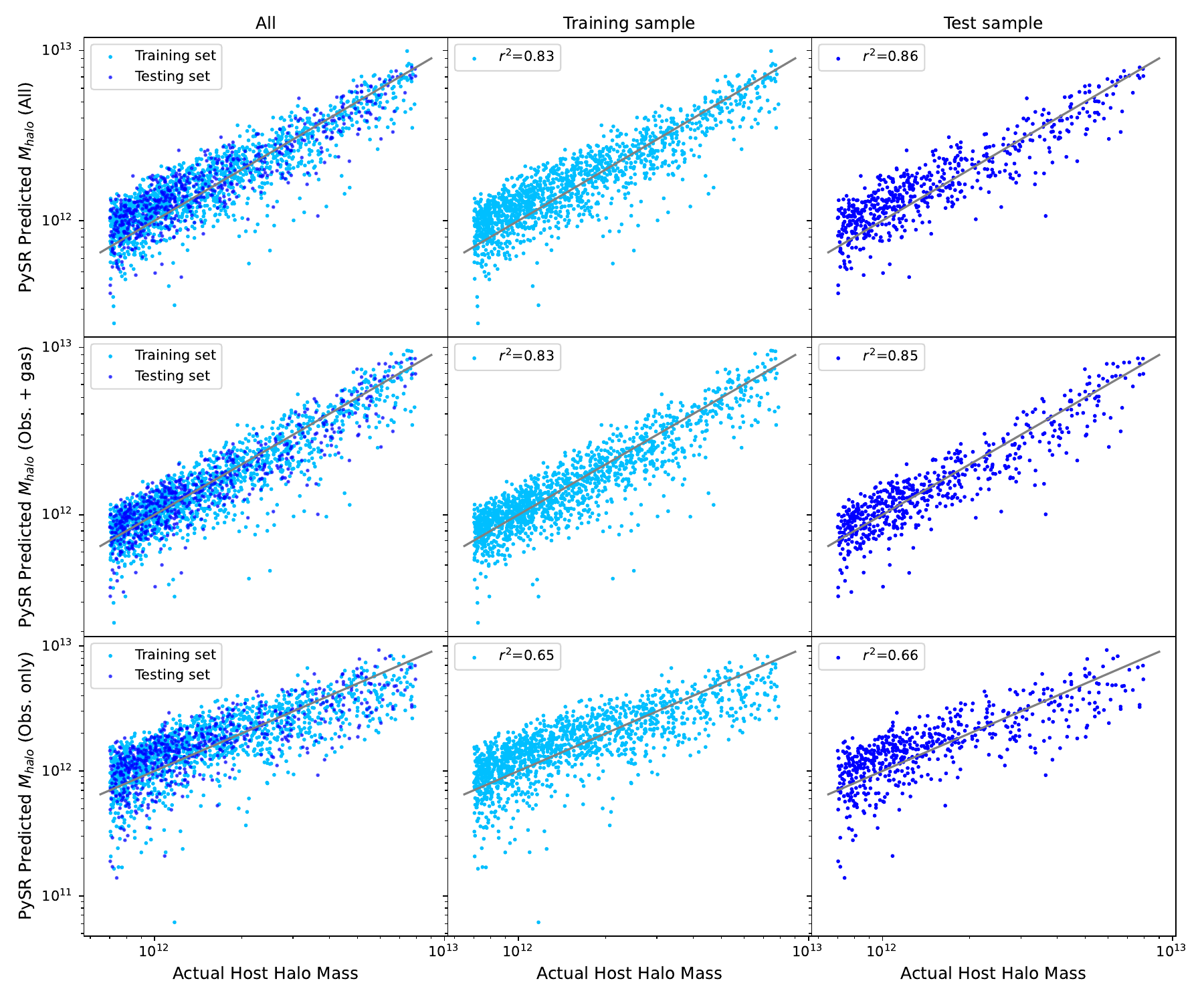}
    \caption{\footnotesize The PySR predicted mass for each set of variables for field halos.  The top row is with all the selected parameters, the middle row is for observable parameters only, and the bottom row is for observables + gas mass.}
    \label{fig:pysr_fields}
\end{figure*}

\begin{figure*}
    \centering
    \includegraphics[width=\linewidth]{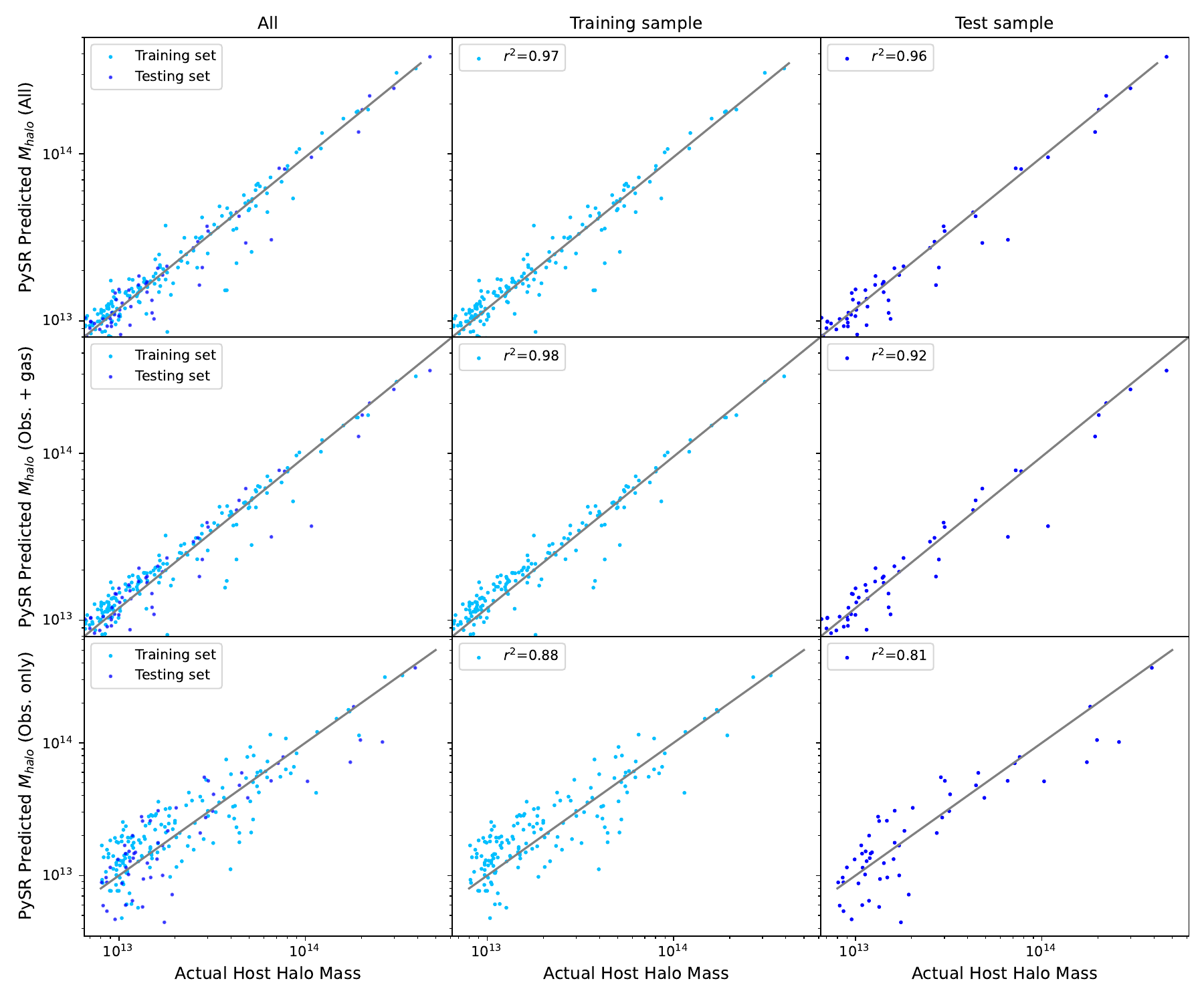}
    \caption{\footnotesize The PySR predicted mass for each set of variables for group+cluster halos.  The top row is with all the selected parameters, the middle row is for observable parameters only, and the bottom row is for observables + gas mass.}
    \label{fig:pysr_gc}
\end{figure*}

We further cross-check this result with PySR, where the formulas infer the scaling relations of the parameters, with their feature importance roughly depending on how often they appear in the outputted formulas.  The results of the feature importances in both random forest and PySR are summarized in Tables \ref{tab:importances} and \ref{tab:eqns}, and the exact halo mass predictions are given in Figures \ref{fig:pysr_fields} and \ref{fig:pysr_gc}.  

\begin{table*}[h]
    \centering
    \begin{tabular}{|p{0.08\textwidth}|p{0.12\textwidth}|p{0.34\textwidth}|p{0.34\textwidth}|}
    \hline  
    Category & Training Set & Random Forest Variables & PySR Variables \\
    \hline 
    Fields & All & gas mass $>$ stellar mass, gr color $>$ sSFR, $r-$band magnitude $>$ half-stellar radius, formation redshift & gas mass $>$ stellar mass $>$ gr color $>$ sSFR $>$ half-stellar radius, $r-$band magnitude $>$ formation redshift \\
    \hline
    Fields & Observables + gas mass & gas mass $>$ stellar mass $>$ gr color $>$ sSFR $>$ half-stellar radius, $r-$band magnitude & gas mass, stellar mass, gr color $>$ sSFR, half-stellar radius, $r-$band magnitude \\
    \hline
    Fields & Observables Only & stellar mass $>$ half-stellar radius, $r-$band magnitude, gr color $>$ sSFR & stellar mass $>$ half-stellar radius $>$ sSFR, gr color $>$ $r-$band magnitude  \\
    \hline
    Groups and clusters & All & gas mass $>$ sSFR $>$ stellar mass $>$ $r-$band magnitude $>$ half-stellar radius, gr color, formation redshift & gas mass $>$ stellar mass, $r-$band magnitude $>$ sSFR, half-stellar radius, gr color, formation redshift \\
    \hline
    Groups and clusters & Observables + gas mass & gas mass $>$ $r-$band magnitude $>$ half-stellar radius, stellar mass $>$ gr color, sSFR & gas mass $>$ sSFR $>$ half-stellar radius $>$ $r-$band magnitude $>$ gr color $>$ stellar mass \\
    \hline
    Groups and clusters & Observables only & stellar mass, $r-$band magnitude $>$ half-stellar radius $>$ gr color, sSFR & stellar mass $>$ half-stellar radius $>$ sSFR, $r-$band magnitude, gr color \\
    \hline
    \end{tabular}
    \caption{\small Training parameters of the central galaxies to find the host halo mass.}
    \label{tab:importances}
\end{table*}

For both runs, the random forest and PySR showed that gas mass, stellar mass, $g-r$ color, and $r-$band magnitude were the important parameters that we focused on, with sSFR and half-stellar radius also having some importance (though not as much as the parameters above).  Since $g-r$ color is connected to the $r-$band magnitude, it is possible that one is pulling mutual significance away from the other, so both quantities are likely correlated to the host halo mass to a non-negligible extent.

There were no noticeable differences between including and not including the formation redshift in the training in both the group+cluster and field halos.  This is likely because the formation redshift seems to have low importance in the random forest algorithm for group and field mass halos (see panel a) in Figures \ref{fig:field_rf} and \ref{fig:gc_rf}).  What did make a difference is taking out the gas mass from the observables - once gas mass is removed, stellar mass became significantly more important, especially in field galaxies.  This is consistent with previous results, where the stellar mass fraction 
\beq
f_{star}=\frac{M_{\ast}}{M_{halo}}
\eeq
peaks at around $10^{12} \text{M}_{\odot}$ halos before decreasing with total halo mass \citep{girelli2020}. 

Along with the parameter significance analysis, we also present some empirical mass predictions with PySR, OLS, and random forest regression.  These results are summarized in Table \ref{tab:goodness_of_fit}.  For the same training set, random forest, OLS, and PySR perform comparably well, with PySR having $r^2$ values 0.02-0.05 higher.  However, random forest regression is more of a black box, whereas OLS and PySR provide analytical, human-interpretable results, which have higher transferability.

Figures \ref{fig:pysr_fields} and \ref{fig:pysr_gc} show how well the PySR function predicts the host halo mass compared to the actual halo mass.  Table \ref{tab:eqns} shows the equations from PySR used to generate Figures \ref{fig:pysr_fields} and \ref{fig:pysr_gc}.  The average values and standard deviations for each prediction set are outlined in Figures \ref{fig:fields_hist} and \ref{fig:gc_hist} in Sec.~\ref{app:results}.

\begin{figure}
    \centering
    \includegraphics[width=0.95\linewidth]{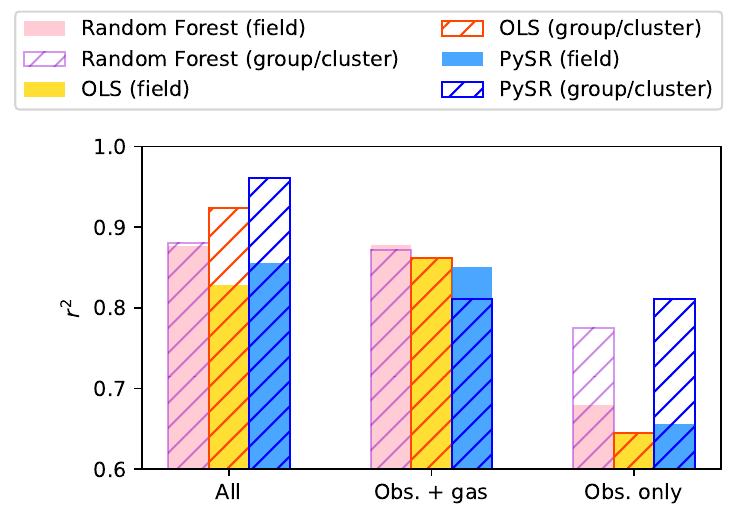}
    \caption{\footnotesize The $r^2$ values of the predicted field halo masses across all methods - random forest, OLS, and PySR - across all parameter sets (all, observables + gas mass, and observables only). }
    \label{fig:r2_bar_summary}
\end{figure}

In addition to the more complex mathematical predictions given by PySR, we use OLS to obtain a linear mathematical form between the halo mass and galaxy observables (as opposed to many possible combinations of linear or nonlinear mathematical functions like PySR).  What is consistent across PySR, OLS, and random forest is that the more parameters used, the better the halo mass prediction does, with the addition of gas mass yielding the largest improvement.  All training sets obtain a better prediction than purely stellar halo mass - see Table \ref{tab:goodness_of_fit} and Figure \ref{fig:r2_bar_summary}.  Overall though, PySR still has a slight edge over the other methods in terms of $r^2$ values, indicating that the relation between host halo mass and galaxy properties is likely nonlinear.

\begin{table*}[h]
    \centering
    \begin{tabular}{|p{0.14\textwidth}|p{0.18\textwidth}|p{0.6\textwidth}|}
    \hline  
    Category & Training Set & Equation \\
    \hline 
    Field & All parameters & $-0.644M_{\ast}^2 - 5.0M_{\ast}x_{\mathrm{sSFR}}x_{gr}(x_{gr} + 1.74) + M_{\ast} + M_{\mathrm{gas}}$  \\
    \hline 
    Field & Observables with gas & $M_{\mathrm{gas}} + \cfrac{M_{\ast} x_{gr}^3 (M_{\mathrm{gas}} + 1.68x_{r\mathrm{band}})}{(M_{\ast} + x_{gr})(x_{\mathrm{rad}} + x_{r\mathrm{band}})(x_{\mathrm{sSFR}} + x_{r\mathrm{band}})}$  \\
    \hline  
    Field & Observables & $1.78M_{\ast}/x_{r\mathrm{band}}^2$  \\
    \hline 
    Group + cluster & All parameters & $-0.0467M_{\mathrm{gas}}(M_{\mathrm{gas}} + log(M_{\ast}))\times(M_{\mathrm{gas}} + x_{r\mathrm{band}} + x_{gr}) + M_{\mathrm{gas}} + 0.07$  \\
    \hline  
    Group + cluster & Observables with gas & $M_{\mathrm{gas}}log(x_{\mathrm{sSFR}} + 2.4) + \cfrac{0.0183}{(-0.93M_{\ast}x_{\mathrm{sSFR}}(66.1M_{\ast}^2 + 76M_{\ast}) + 1.3)}$  \\
    \hline 
    Group + cluster & Observables & $0.86M_{\ast}(x_{\mathrm{rad}} + 0.564)$  \\
    \hline 
    \end{tabular}
    \caption{The variables are as follows: $M_{\ast}$ is stellar mass, $M_{\mathrm{gas}}$ is gas mass, $x_{\mathrm{rad}}$ is half stellar radius, $x_{\mathrm{sSFR}}$ is the sSFR, $x_{z}$ is the formation redshift, $x_{r\mathrm{band}}$ is the $r-$band magnitude, and $x_{gr}$ is the $g-r$ color.  They are from the central galaxies of the halos.}
    \label{tab:eqns}
\end{table*}

\begin{table*}[h]
    \centering
    \begin{tabular}{|c|c|c|c|}
    \hline  
    Category & Galaxy Properties & Training Algorithm & $R^2$ \\
    \hline 
    Field & All parameters & Random Forest &  0.88 \\
    \hline 
    Field & Observables only & Random Forest & 0.68  \\
    \hline 
    Field & Observables with gas & Random Forest & 0.88 \\
    \hline
    Field & All parameters & Least Squares &  0.85 \\
    \hline 
    Field & Observables only & Least Squares & 0.65 \\
    \hline 
    Field & Observables with gas & Least Squares & 0.85 \\
    \hline 
    Field & Stellar Mass only & Least Squares & 0.64 \\
    \hline 
    Field & All parameters & PySR & 0.84 \\
    \hline 
    Field & Observables only & PySR & 0.67 \\
    \hline 
    Field & Observables with gas & PySR & 0.84 \\
    \hline 
    Group + cluster & All parameters & Random Forest & 0.85 \\
    \hline 
    Group + cluster & Observables only & Random Forest & 0.77 \\
    \hline 
    Group + cluster & Observables with gas & Random Forest & 0.84 \\
    \hline 
    Group + cluster & All parameters & Least Squares & 0.96 \\
    \hline 
    Group + cluster & Observables only & Least Squares & 0.83 \\
    \hline 
    Group + cluster & Observables with gas & Least Squares & 0.96 \\
    \hline 
    Group + cluster & Stellar Mass only & Least Squares & 0.81 \\
    \hline 
    Group + cluster & All parameters & PySR & 0.97 \\
    \hline 
    Group + cluster & Observables only & PySR & 0.85 \\
    \hline 
    Group + cluster & Observables with gas & PySR & 0.96 \\
    \hline
    \end{tabular}
    \caption{The $R^2$ values of each category of galaxy observables using a different method of host halo mass prediction. }
    \label{tab:goodness_of_fit}
\end{table*}

From the results above, it can be seen that gas mass is the most important quantity in determining the underlying host halo mass, followed by stellar mass, color, and magnitude; see Table \ref{tab:importances} for a summary.  It should also be noted that stellar mass, color, and magnitude importances increase once we take out gas mass as a parameter.  Formation redshift, sSFR, and half-stellar radius comparatively do not have a high importance - this is likely because the sSFR is connected to both stellar mass and color, and the half-stellar radius does not encompass the dark matter on the outskirts of the galaxies, making it less correlated with the overall dark matter content of the halo.  Across all methods, adding in more parameters improves the host halo mass prediction - see Tables \ref{tab:goodness_of_fit} and \ref{tab:goodness_of_fit_sub}.  This shows that halo mass likely has a multi-dimensional relation with the properties of its central galaxy beyond the stellar mass.

\subsection{Subhalo Mass for Satellites}
\label{subsec:results_subhalos}

\begin{figure}
    \centering    \includegraphics[width=0.85\linewidth]{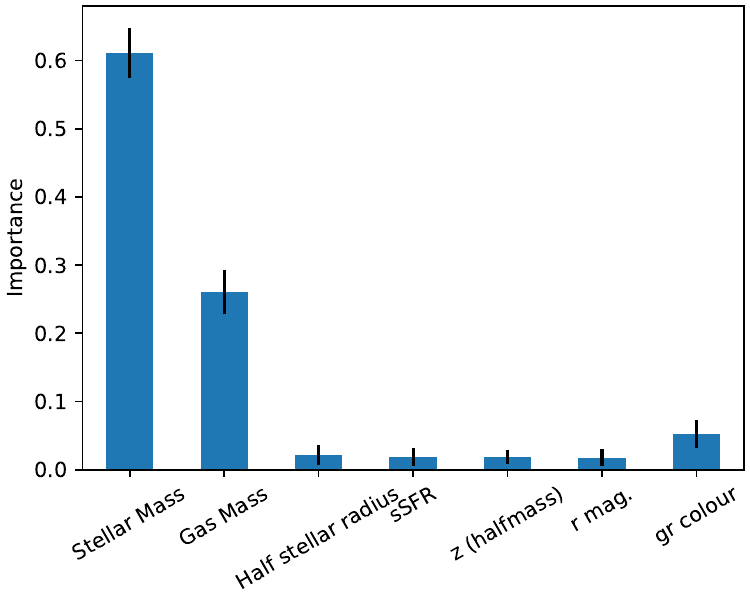} \\
    {\footnotesize a) All features that could be important for field satellite subhalos.}
        \vspace{2mm}    \includegraphics[width=0.85\linewidth]{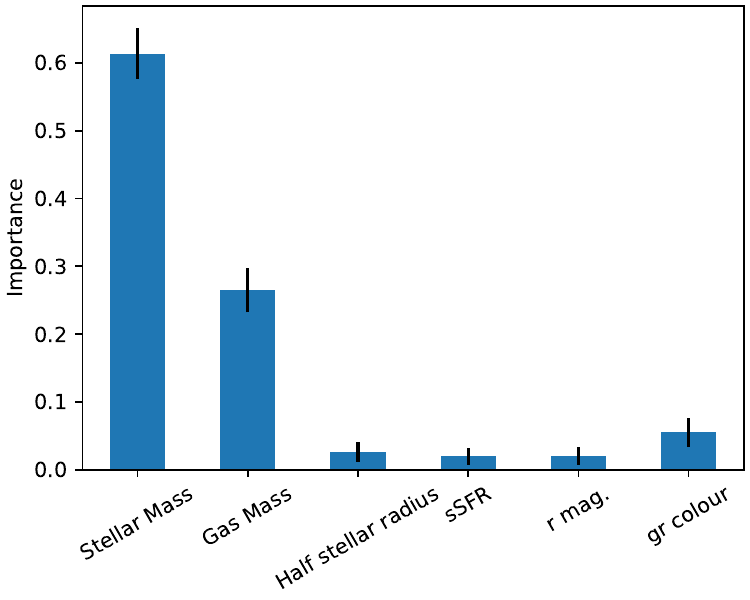} \\
    {\footnotesize b) Features that are found in observations + gas mass.}
        \vspace{2mm}    \includegraphics[width=0.85\linewidth]{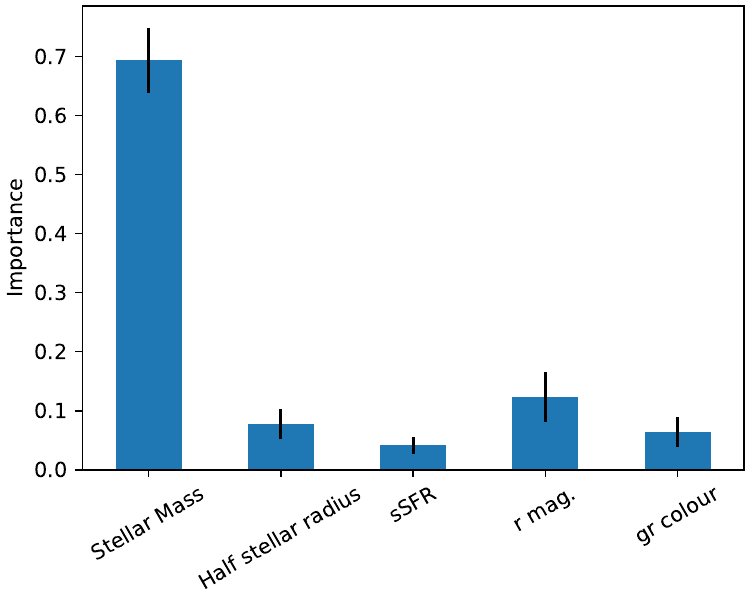} \\
         {\footnotesize c) Features that are observable in real space.}
    \caption{\footnotesize Random forest feature importance of different combinations of satellite galaxy features correlated with their subhalo masses in field-sized host halos.}
    \label{fig:field_sub_rf}
\end{figure}

In addition to estimating host halo masses using the central galaxies, we also used the same process to estimate the subhalo masses of satellite galaxies.  Similar to centrals and host halos, we start with the random forest analysis to estimate feature importance.  We use the same parameter sets in the list \ref{list:properties} for both centrals and satellite galaxies.  We found that in contrast to centrals and host halos, the importance of formation stellar mass is actually higher than gas mass in field subhalos - see Figure \ref{fig:field_sub_rf} and Table \ref{tab:importances_sub}.  The importance of a satellite's $g-r$ color also is higher than that of its $r-$band magnitude, suggesting that field satellites have a different relation to its subhalos than the field centrals do with their host halos.

\begin{figure}
    \centering    \includegraphics[width=0.85\linewidth]{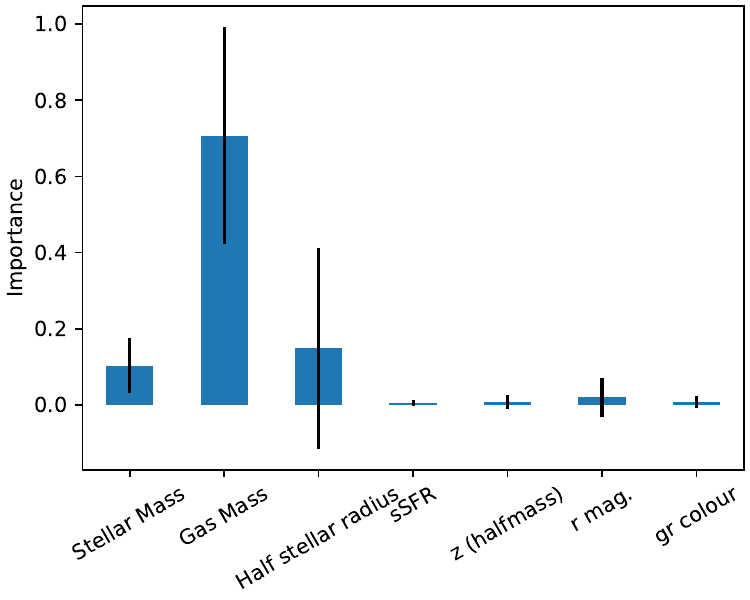}
         \\ {\footnotesize a) All training parameters for satellite galaxies of group+cluster mass range.}
         \vspace{2mm}
    \includegraphics[width=0.85\linewidth]{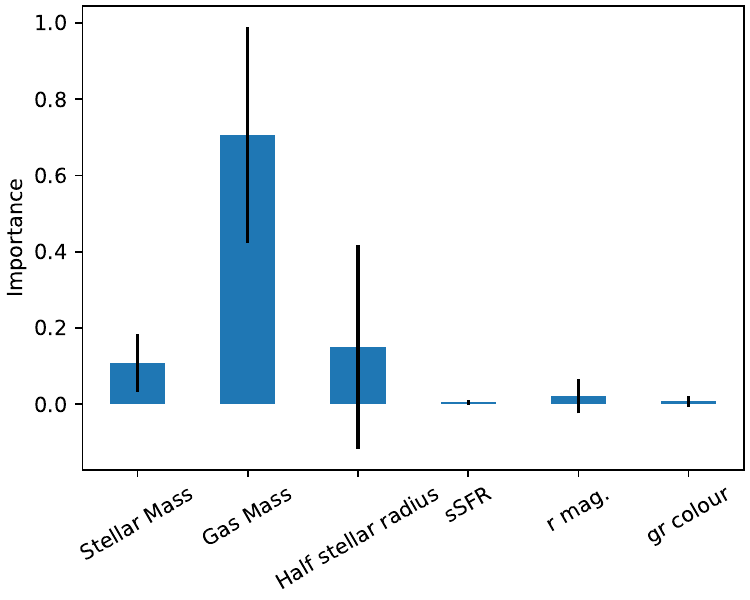}
         \\ {\footnotesize b) Features that are found in simulations.}
         \vspace{2mm}
    \includegraphics[width=0.85\linewidth]{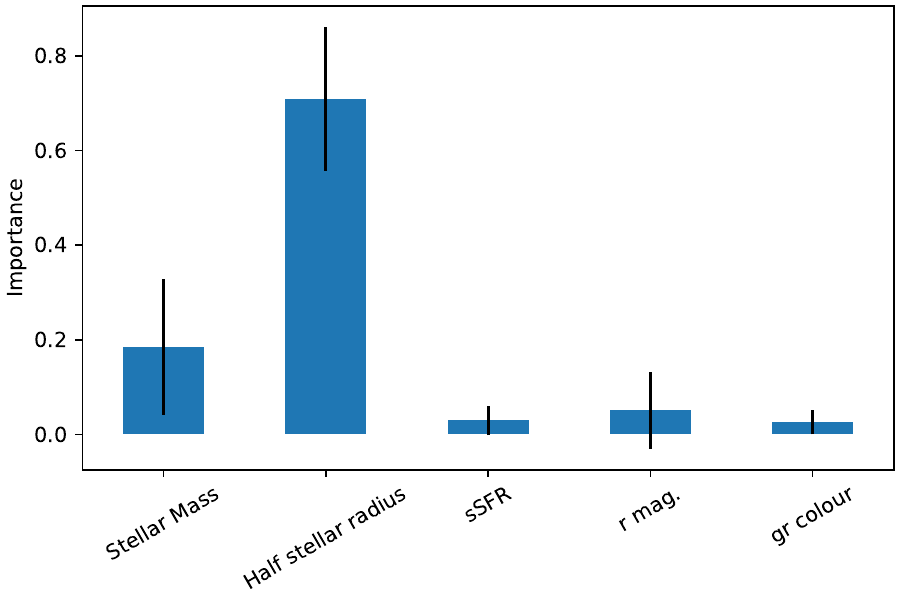}
         \\ {\footnotesize c) Features that are observable in real space.}
         \vspace{1mm}
        \caption{\footnotesize Random forest feature importance of different combinations of satellite galaxy features in group and cluster-sized host halos (if the group and cluster samples are combined).}
    \label{fig:gc_sub_rf}
\end{figure}

\begin{figure*}
    \centering    \includegraphics[width=0.9\linewidth]{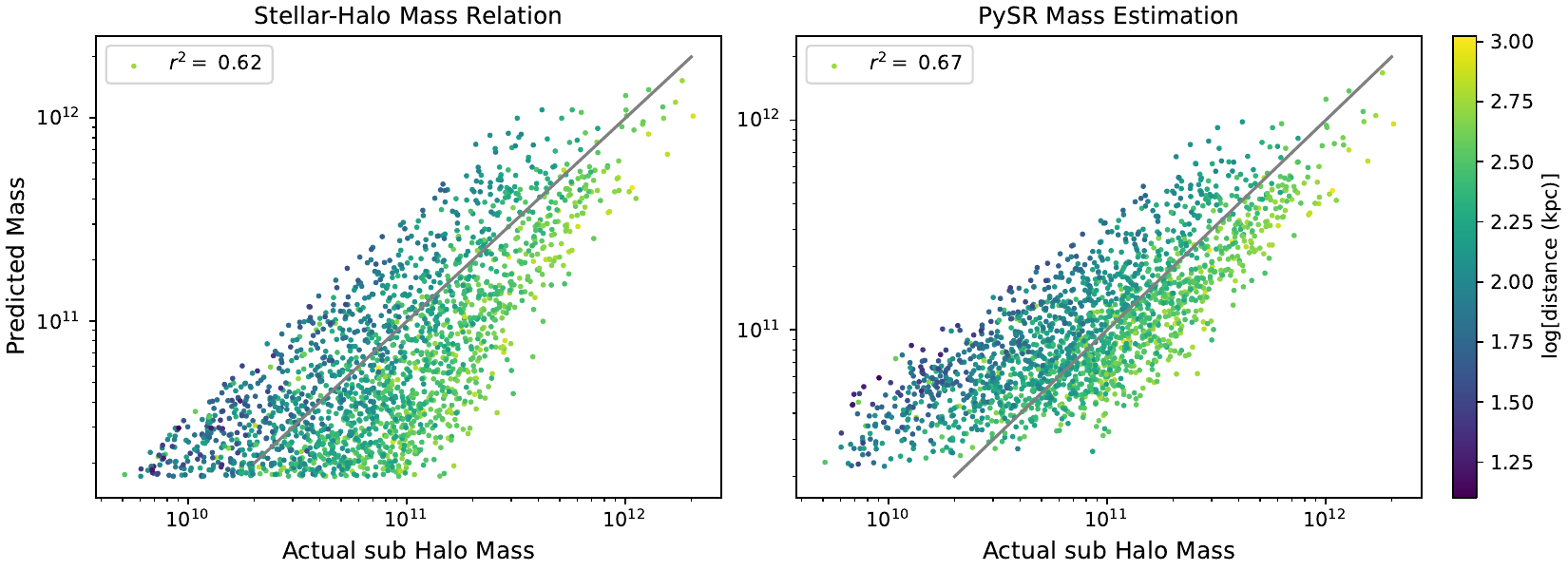}
    \caption{\footnotesize Prediction of subhalo masses of our analysis compared to stellar-halo mass scalings, for satellites residing in field mass halos.  From the $r^2$ value, it can be seen that the machine learning analysis beyond using only stellar mass does yield an improvement.}
    \label{fig:subhalo_mass_fields}
\end{figure*}

\begin{figure*}
    \centering    \includegraphics[width=0.9\linewidth]{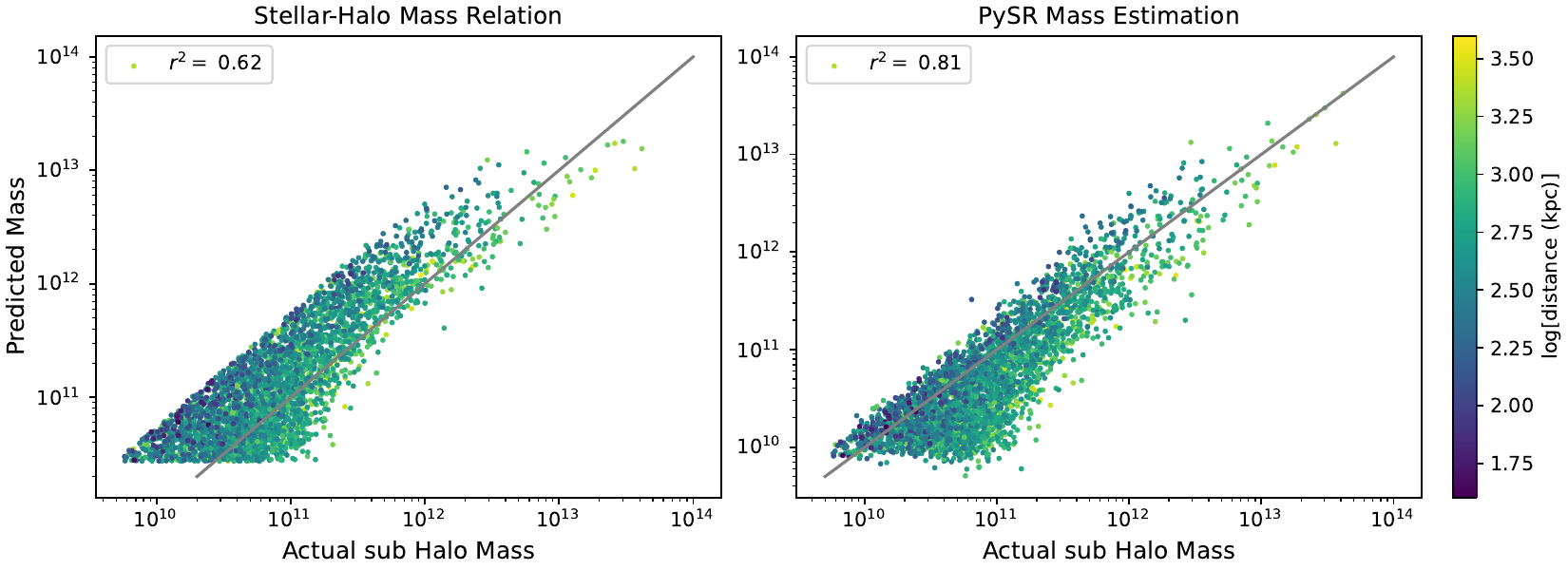}
    \caption{\footnotesize Prediction of subhalo masses of our analysis compared to stellar-halo mass scalings, for satellites residing in group and cluster mass halos.  Again from the $r^2$ value, it can be seen that the machine learning analysis does improve results.}
    \label{fig:subhalo_mass_gc}
\end{figure*}

The feature importance results for satellites and subhalos in groups and clusters were more similar to their central feature results.  Like their central counterparts, satellites in groups and clusters also have gas mass as their most important feature - see Figure \ref{fig:gc_sub_rf} and again, Table \ref{tab:importances_sub}.  Compared to field satellites, satellites in groups and clusters seem to have overall less deviation when it comes to their feature importance and subhalo mass.  In both field halos and groups/clusters though, it can be seen that a non-linear, multi-dimensional estimates of subhalo mass obtain better predictions than standard stellar-to-halo mass relations, particularly for subhalos residing in groups and clusters - see Figures \ref{fig:subhalo_mass_fields}, \ref{fig:subhalo_mass_gc}, and Table \ref{tab:goodness_of_fit_sub}.

\begin{table*}[ht!]
    \centering
    \begin{tabular}{|p{0.08\textwidth}|p{0.12\textwidth}|p{0.34\textwidth}|p{0.34\textwidth}|}
    \hline  
    Category & Training Set & Random Forest Variables & PySR Variables \\
    \hline 
    Fields & All & gas mass $>$ stellar mass, gr color $>$ sSFR, $r-$band magnitude $>$ half-stellar radius, formation redshift & gas mass $>$ stellar mass $>$ gr color $>$ sSFR $>$ half-stellar radius, $r-$band magnitude $>$ formation redshift \\
    \hline
    Fields & Observables Only & stellar mass $>$ half-stellar radius, $r-$band magnitude, gr color $>$ sSFR & stellar mass $>$ half-stellar radius $>$ sSFR, gr color $>$ $r-$band magnitude  \\
    \hline
    Groups and clusters & All & gas mass $>$ sSFR $>$ stellar mass $>$ $r-$band magnitude $>$ half-stellar radius, gr color, formation redshift & gas mass $>$ stellar mass, $r-$band magnitude $>$ sSFR, half-stellar radius, gr color, formation redshift \\
    \hline
    Groups and clusters & Observables only & stellar mass, $r-$band magnitude $>$ half-stellar radius $>$ gr color, sSFR & stellar mass $>$ half-stellar radius $>$ sSFR, $r-$band magnitude, gr color \\
    \hline
    \end{tabular}
    \caption{\small Feature importances for subhalo parameter training sets.}
    \label{tab:importances_sub}
\end{table*}

\begin{table*}[ht!]
    \centering
    \begin{tabular}{|c|c|c|c|}
    \hline  
    Category & Galaxy Properties & Training Algorithm & $R^2$ \\
    \hline 
    Field & All parameters & Random Forest & 0.88 \\
    \hline 
    Field & Observables only & Random Forest & 0.60 \\
    \hline 
    Field & All parameters & Least Squares & 0.67 \\
    \hline 
    Field & Observables only & Least Squares & 0.65 \\
    \hline 
    Field & Stellar Mass only & Least Squares & 0.59 \\
    \hline
    Field & Observables only & PySR & 0.67 \\
    \hline 
    Group + cluster & All parameters & Random Forest & 0.67 \\
    \hline 
    Group + cluster & Observables only & Random Forest & 0.54 \\
    \hline 
    Group + cluster & All parameters & Least Squares & 0.63 \\
    \hline 
    Group + cluster & Observables only & Least Squares & 0.60 \\
    \hline 
    Group + cluster & Stellar Mass only & Least Squares & 0.55 \\
    \hline
    Group + cluster & Observables only & PySR & 0.71 \\
    \hline
    \end{tabular}
    \caption{The $r^2$ values of how well satellite galaxy properties can predict subhalo mass.}
    \label{tab:goodness_of_fit_sub}
\end{table*} 

\section{Discussion}
\label{sec:discussion}

\subsection{Performance of Predicted Halo Mass vs True Halo Mass from TNG100}\label{subsec:halo_prediction_accuracy}

From Figures \ref{fig:pysr_fields}-\ref{fig:pysr_gc}, it can be seen that while the functions in Table \ref{tab:eqns} is able to fit the actual halo mass trend, there is a large scatter around a line of perfect fit (the grey line in Figures \ref{fig:pysr_fields}-\ref{fig:pysr_gc}).  This scatter is likely caused by factors such as scatter in the original host masses, environmental and assembly histories, and potential correlations between the parameters themselves (stellar mass and sSFR, $r-$band magnitude and $g-r$ color).  Environmental factors such as feedback effects, mergers, and interactions can cause intrinsic scatter in the original data points.  How to quantify and constrain the effects of this scatter will be useful for understanding halo mass evolution, but that is beyond the scope of this paper.

As for the correlations in the observable properties themselves, there is also some degeneracy in their impact on the halo mass.  Properties such as color and star formation rate measurements affect each other nonlinearly, which would affect their ``independent" feature importance on the halo mass.  Future work studying specific mechanisms for sSFR and color would be important for disentangling these degeneracies, although that is beyond the scope this work as well.

\subsection{Performance of Different Parameter Sets and Different Halo Masses}

The intrinsic scatter from the original data is expected to impact the final predictive formulas from PySR, which we explored in Sec. \ref{app:results}.  From Figures \ref{fig:fields_hist}-\ref{fig:gc_hist} and Table \ref{tab:sigma}, it can be seen that taking training sets of different parameters does not vastly alter the parametric forms given in the symbolic regression fitting formula.  
This is likely because PySR is designed to always fit the best form regardless of how many parameters are given.

The gas mass is one of the most important parameters, which is consistent with theoretical expectations, since gas mass is expected to make up a significant fraction of the halo mass.  Formation redshift, in contrast, seems to be rather unimportant in predicting the total host halo mass for field and group mass halos.  This shows that the present day host halo mass (for halos $\leq 10^{14}M_{\odot}$) is weakly correlated to their mass assembly history.  Despite the gas mass having high significance, we found that adding in other galaxy parameters such as the sSFR, color, magnitude, and half-stellar radius can get a tighter constraint on the estimated halo mass (see Table \ref{tab:goodness_of_fit})

An additionally interesting result is that the importance of gas mass with respect to the halo mass decreases with the most massive halos.  While there were only 17 clusters in this analysis and this is not a statistically significant result, it is a potentially interesting finding that is worth exploring in larger, high-resolution simulations as they become available.  A possible explanation for this is that centrals in clusters formed much of their stars a long time ago, so most of their cold gas mass has been exhausted - gas becomes a comparatively smaller fraction of their total mass.  Another interesting characteristic about these 17 clusters is how much formation redshift is correlated with the overall host halo mass; it is notably more significant in the most massive halos (clusters) - see Figure \ref{fig:cluster_rf} - compared to the lower mass groups (or field) - see Figures \ref{fig:field_rf} and \ref{fig:group_rf}.  However, again, given the sample size, this result is not yet statistically significant - it could be worth exploring in simulations focused specifically on clusters.  

From Figures \ref{fig:pysr_fields}-\ref{fig:pysr_gc} it looks like group+cluster halo masses are better predicted than those of field halos due their $r^2$ value being closer to 1.0.  In particular, the $r^2$ value of groups+clusters halo mass predictions is 0.85, 0.18 larger than the field prediction of 0.67 for observables only.  This shows that for groups and clusters, the galaxy prediction captures their variation better.  However, Table \ref{tab:sigma} and Figure \ref{fig:violin} show that the average ratio of $\frac{\mathrm{predicted}\ M_{\mathrm{halo}}}{\mathrm{true}\ M_{\mathrm{halo}}}$ for field halos is around 1.0, compared to 1.3 for groups and clusters.  The difference between their average ratio of $\frac{\mathrm{predicted}\ M_{\mathrm{halo}}}{\mathrm{true}\ M_{\mathrm{halo}}}$ is also over 5$\sigma$, which is quite significant.  

Field mass predictions likely have more scatter because field halos have a larger diversity in galaxy properties compared to higher mass halos.  This diversity introduces more intrinsic scatter in field halos, which results in a lower $r^2$ value, but their large sample size make their $\frac{\mathrm{predicted}\ M_{\mathrm{halo}}}{\mathrm{true}\ M_{\mathrm{halo}}}$ ratio closer to 1.0.  In contrast, groups and clusters have lower intrinsic scatter, which results in a higher $r^2$ value, but due to their smaller sample size, the $\frac{\mathrm{predicted}\ M_{\mathrm{halo}}}{\mathrm{true}\ M_{\mathrm{halo}}}$ deviation from 1.0 is more likely.

\subsection{Predictions with Observational Data for Groups and Clusters}

\begin{figure*}
    \centering
    \includegraphics[trim=0 0 0 15,clip,width=0.9\linewidth]{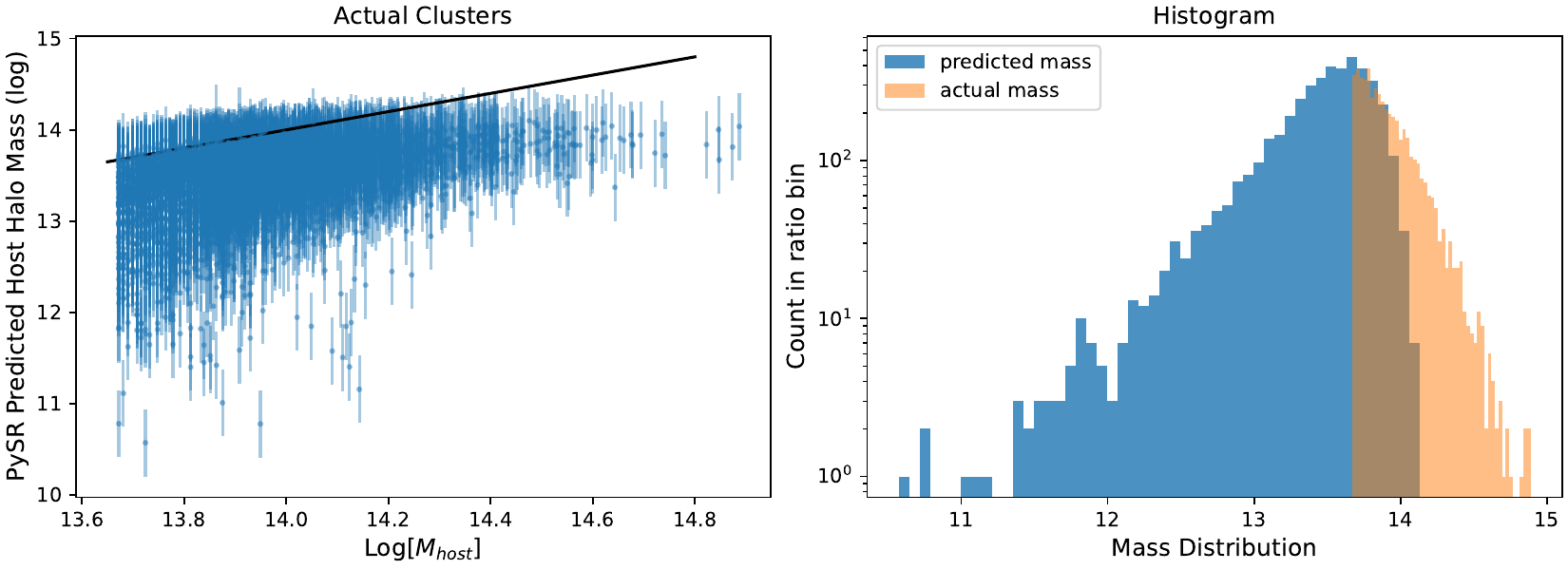}
    \caption{\small Left panel: The predicted mass of our model vs the predicted mass from \citet{2021ApJ...909..143Y}.  Right panel: histogram of our halo mass predictions vs reported halo mass predictions from \citet{2021ApJ...909..143Y}.}
    \label{fig:yang_obs}
\end{figure*}

The goal of this analysis is to be able to apply a model of dark matter halo mass to observations.  We take the model we have trained so far on observable galaxy properties from simulations and test it on observational survey data.  To test the performance of our model, we compare it to some halo mass estimates from observations that already exist in the literature.  These halos have mass measurements using weak lensing, X-ray, and different scaling relations \citep[e.g.,][]{2021ApJ...909..143Y, Herbonnet:2019byy, Li:2024ogc}.  We show the results of our estimated comparison to the data in \citet{2021ApJ...909..143Y} in Figure \ref{fig:yang_obs}.

From Figure \ref{fig:yang_obs}, for a fraction of the sample, our predictions are consistent with the observed measurements.  For the rest of the sample, we can see that our training method on TNG100 consistently underpredicts the halo mass of the central galaxies compared to the \citet{2021ApJ...909..143Y} results.  To investigate whether this difference is the result of our predictive model or the galaxy property distribution itself, we compare the simulation galaxy properties to the observed properties in Figure \ref{fig:obs_sims} in the Appendix \ref{app:scatter}.  We also include the EAGLE simulation \citep{2015MNRAS.446..521S}, specifically the RefL0100N1504 run, which has a comparable comoving volume to TNG100 (67.8 Mpc/$h$ versus 75 Mpc/$h$) and uses a similar cosmological framework (Planck 2013 \citep{Planck:2013pxb} versus Planck 2015 \citep{Planck:2015fie}).  Since large differences between the color–halo mass and stellar mass distributions were observed between simulations - both TNG and EAGLE - and observations, there is a high possibility that the underprediction in our results is due to differences in the galaxy property distributions.

Despite these offsets, the $r$-band magnitude–stellar mass relation in TNG100 does still match the general trend of the observational data (middle panel of Figure \ref{fig:obs_sims}), and color-based quantities seem to be systematically offset by approximately $\sim 1$ dex (right panel of Figure \ref{fig:obs_sims}).  This could be due to the over-cooling problem in simulations, which leads to a sSFR that is likely higher than what exists in real space.  This would cause the stellar mass in simulated centrals to be larger than what is in observed centrals, which would affect halo mass predictions.

It should be further noted that what is classified as a central galaxy in the observational data might not be a central in real space, introducing an additional source of error for mass estimation.  The observed masses could have uncertainties in their measurements that lead to an over-predicted halo mass as well; currently, it is unclear whether the discrepancy between our prediction and the observed masses are because of our analyses, or potential observational errors themselves.  These sources of uncertainties can explain why there is a visible difference in our predicted halo mass and the observed halo masses.  However, the fraction of overlapping predictions between our method and observations is still promising - this shows that it is possible to constrain halo mass using observable galaxy properties.  Additionally, we also do not compare our predictions to field sized halo masses, since unlike clusters, they are much harder to have mass measurements; they usually have much weaker signals in lensing and there is no publicly available large sample of rotation curve based halo mass measurements.

\subsection{Comparison with Stellar-to-Halo Mass Relation}
\label{subsec:discussion_stellar_mass_comparison}
Previously, one of the most common and accurate ways to infer halo mass was using the stellar-to-halo mass relation from the halo's central galaxy \citep{1976ApJ...203..297S, 2016MNRAS.457.3200M, Bilicki:2021hgn}.  Here, we compare the mass estimations between our multi-dimensional relations and the simple stellar-to-halo mass relation.  From the $R^2$ values in Tables \ref{tab:goodness_of_fit} and \ref{tab:goodness_of_fit_sub}, it can be seen that both the halo and subhalo masses can be better predicted by using more of their galaxy properties than just stellar mass (see the scatter of the data and $r^2$ values in Figures \ref{fig:subhalo_mass_fields} and \ref{fig:subhalo_mass_gc}).  The $R^2$ improves when adding in observables, and does slightly better with a nonlinear, multi-dimensional relation compared to a linear one (PySR vs. OLS).  This finding indicates that for a more accurate host halo mass estimation, we need to go beyond the stellar-to-halo mass relation and adding more observable galaxy properties improves results.  We should also not necessarily assume the relations are linear, as we can see that PySR still does marginally better compared to the purely linear cases.  Other works have proposed that stellar-to-halo mass relations vary across different galaxy types and have dependence on other galaxy properties \citep{2020MNRAS.499.5656R, 2023MNRAS.518.1002R}, which provides additional context for going beyond only using stellar mass when estimating halo masses.

There is also a trend that the subhalo masses are overpredicted by satellites close to the central of the host halo, while they become underpredicted by satellites farther from the central - see Figures \ref{fig:subhalo_mass_fields} and \ref{fig:subhalo_mass_gc}.  This could be due to a particle membership mis-assignment between subhalos and their underlying host halos, though the exact reason behind this is unknown as of this work and would be interesting to explore more in the future.

\subsection{Comparison with Mass Richness in Clusters}
\label{subsec:discussion_mass_richness}
\begin{figure*}[htp!]
    \centering
    \gridline{ \fig{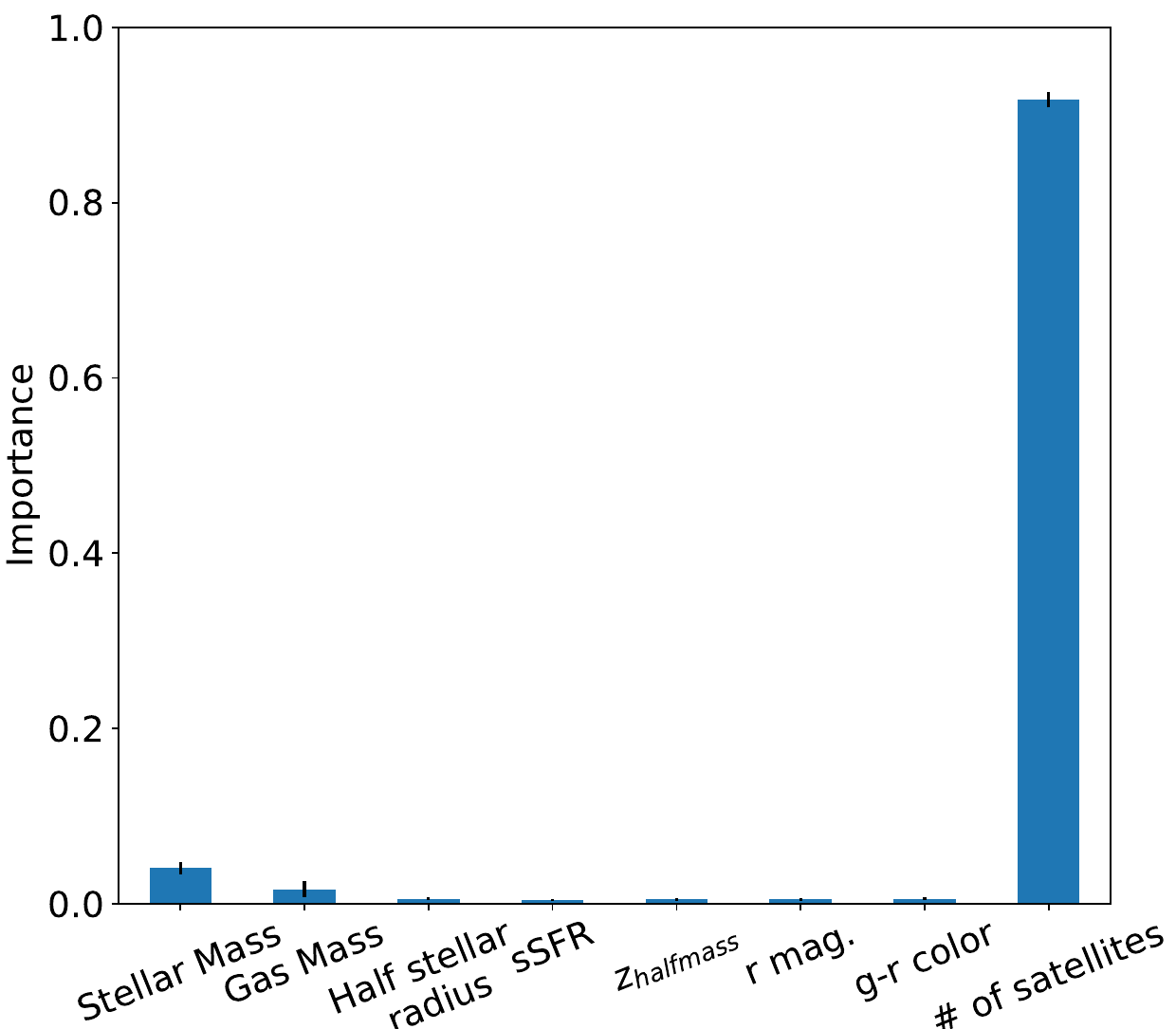}{0.45\textwidth}{\footnotesize a) Field halos.}
    \fig{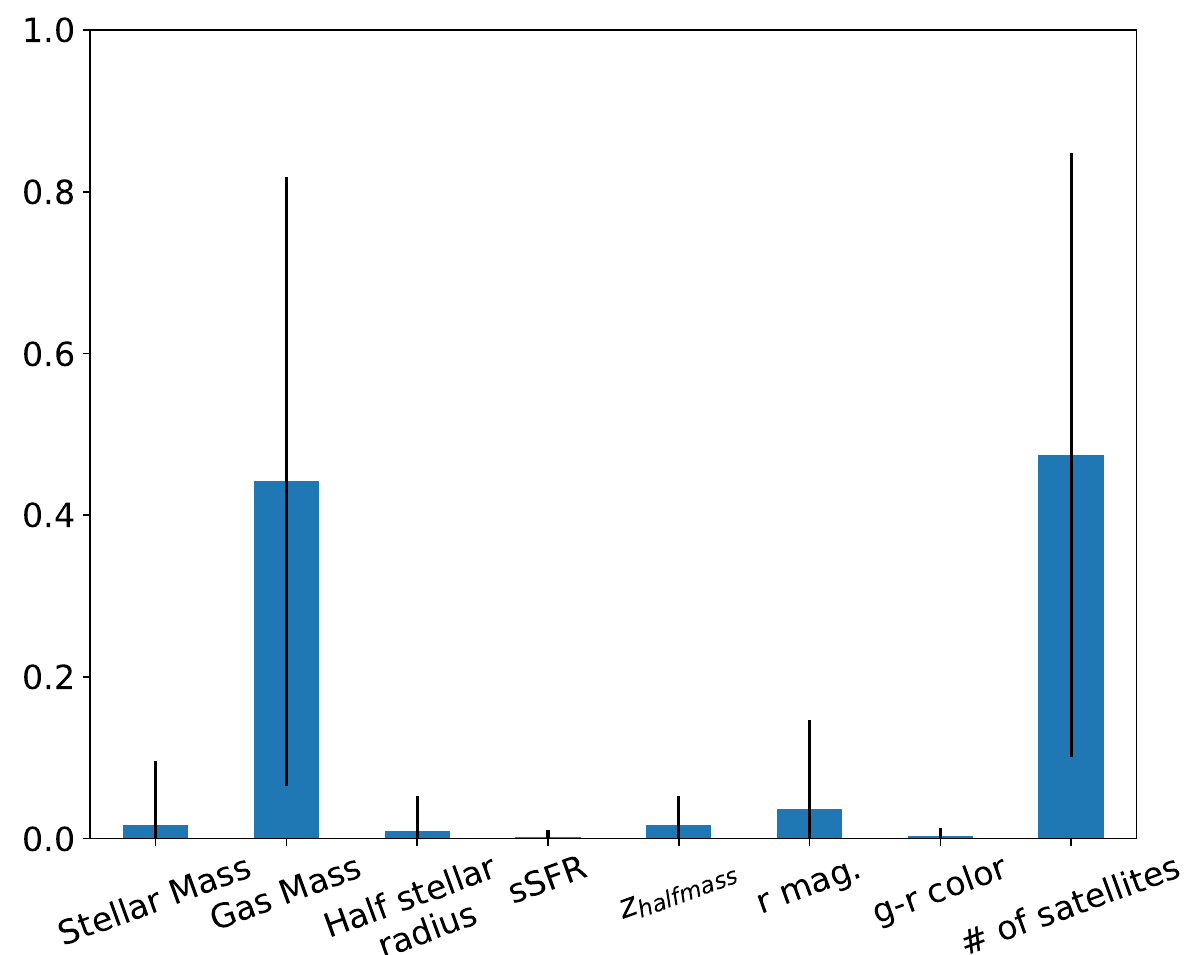}{0.48\textwidth}{(b) Groups and clusters.} }
    \caption{\footnotesize The random forest results for all parameters, including the total number of galaxies (richness) in a halo.  It can be seen that the richness is the feature with the highest importance with respect to the total host halo mass.}
    \label{fig:host_mass_satnum}
\end{figure*}

The results above in Sec. \ref{sec:results} show that the total mass of the host halos scales well with the gas mass of the central galaxies, as well as stellar mass, color, and magnitude fairly well.  One key property that we have not looked at so far is the relation between mass and richness (number of galaxies) in halos.  Particularly in more massive halos such as groups and clusters, previous works \citep{Murata:2019fxk} have found that richness has a tight correlation with the host halo mass.  Figure \ref{fig:host_mass_satnum} shows the random forest feature importance of the number of galaxies inside the host halo alongside the other properties we considered for centrals in groups and clusters.  It can be seen that the number of satellites is by far the most important feature, which indicates that richness would be a strong predictor of the host halo mass.  One note of caution is that the small group/cluster sample size may not fully capture the diversity of the group and cluster populations, but the strong richness importance indicates that adding the number of member galaxies of a halo would improve its mass predictions significantly.  As of now, estimating the number of galaxies in each halo is nontrivial due to observational limits such as apparent magnitude cuts and redshift measurement uncertainties.  An additional source of halo uncertainty is its boundary definition, as there are different ways of defining a halo's ``true" radial boundary; this problem is beyond the scope of this paper.  In future work, it would be interesting to see how richness can be incorporated into a multi-dimensional halo mass prediction model to get tighter constraints.
         
\section{Conclusion and Future Prospects}
\label{sec:conclusion}
In this work, we have presented an empirical relation to predict host halo masses using observable galaxy properties.  We also showed a comparison of how each observable property's correlation strength scales with the host halo mass.  We find that gas mass is the most correlated property with the total host halo mass across all halo masses.  Color and/or magnitude are also decent tracers of halo mass that are directly observable.  Overall, using multiple galaxy properties in the analysis improved results over the standard stellar-to-halo mass relation, which implies that the relation between halo mass and its central galaxy is multi-modal. 

This motivates future work to disentangle the parameters' intrinsic correlations.  If we were able to find the exact relations between properties such as stellar mass and magnitude (or sSFR and color), we could potentially have a better predictive model on how halo mass depends on each observable galaxy property.  It would also be interesting to explore cluster mass halos as a separate category from groups, in simulations with large enough volume to produce a statistically significant number of clusters.  While there were only a few cluster sized halos in our sample, they exhibit different behaviour compared to fields or groups, such as higher formation redshift importance and lower gas mass importance in their centrals.  Whether or not this trend holds as the number of cluster sized halos increases remains to be seen, and it would be an interesting analysis to see how they differ from group sized halos in the same simulation.

\section*{Acknowledgements}
This research was funded by the Natural Sciences and Engineering Research Council of Canada (NSERC) through grants held by AC's supervisor, Niayesh Afshordi. Additional support was provided through the authors' affiliations with University of Waterloo and Perimeter Institute of Theoretical Physics.  The authors gratefully acknowledge this support.


\appendix
\label{sec:appendix}

\section{Additional materials for results}
\label{app:results}

\begin{figure*}[htp!]
     \centering
      \begin{minipage}[t]{0.34\textwidth}
         \centering         \includegraphics[width=\textwidth]{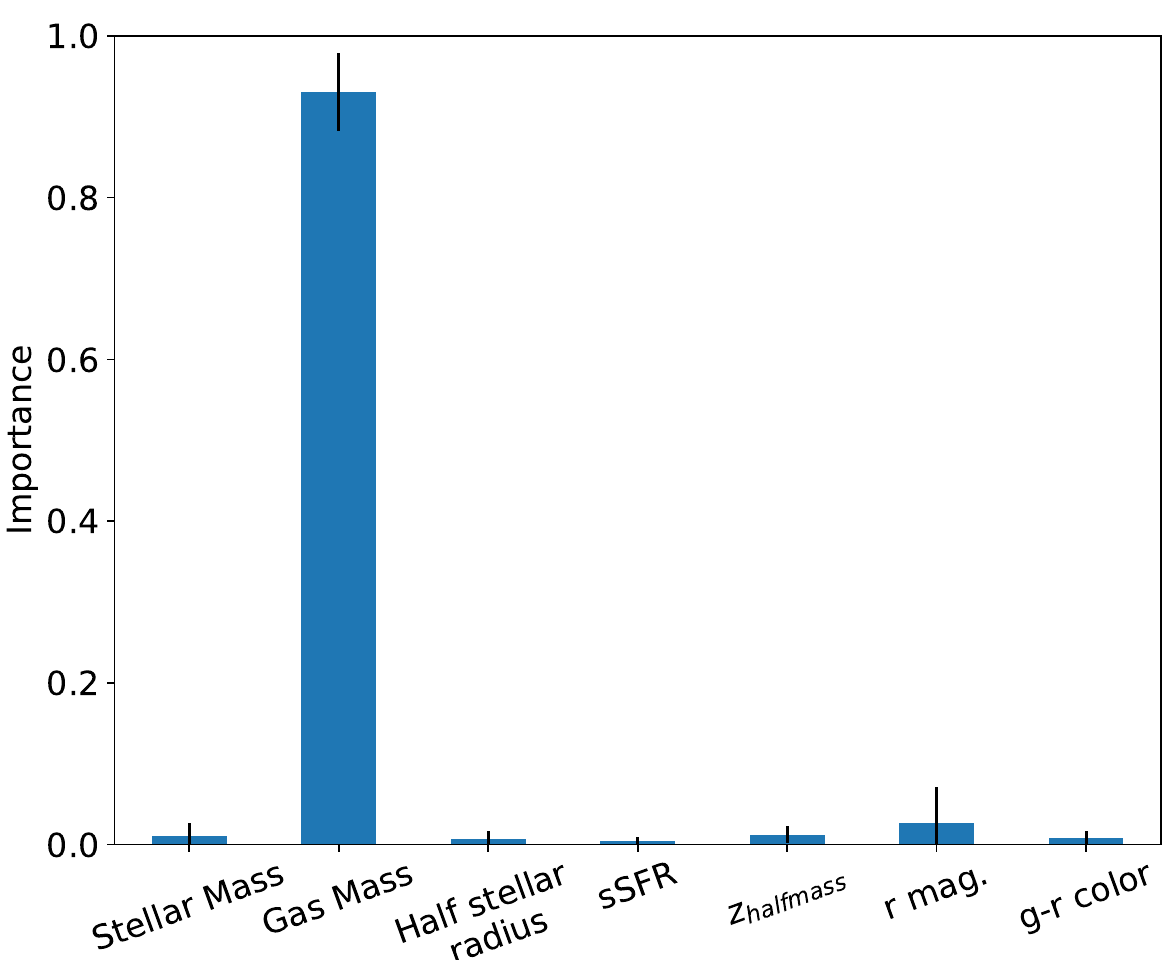}
         \\ \footnotesize All features that could be important for group central galaxies.
         \label{fig:group_rf_all}
      \end{minipage}
      \begin{minipage}[t]{0.31\textwidth}
         \centering         \includegraphics[width=\textwidth]{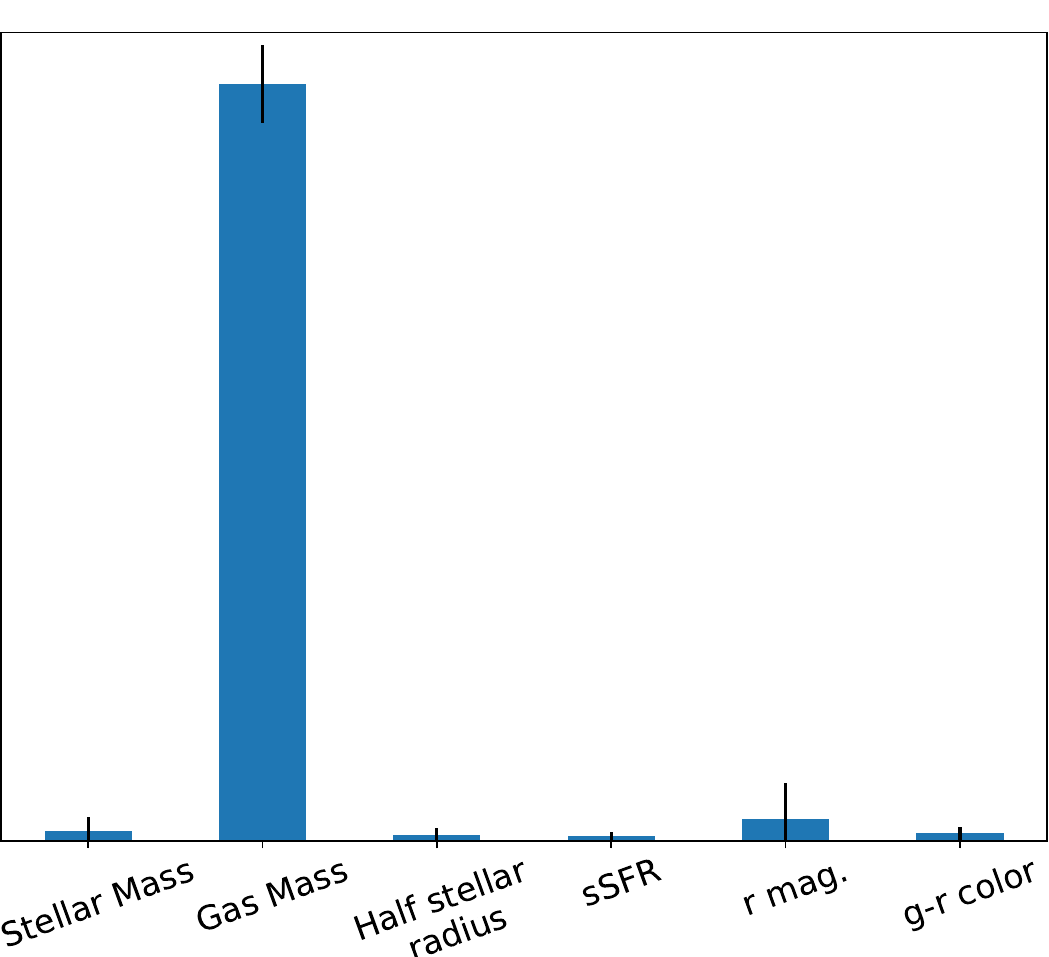}
         \\ \footnotesize Features that are found in simulations.
         \label{fig:group_rf_sim}
      \end{minipage}
      \begin{minipage}[t]{0.31\textwidth}
         \centering         \includegraphics[width=\textwidth]{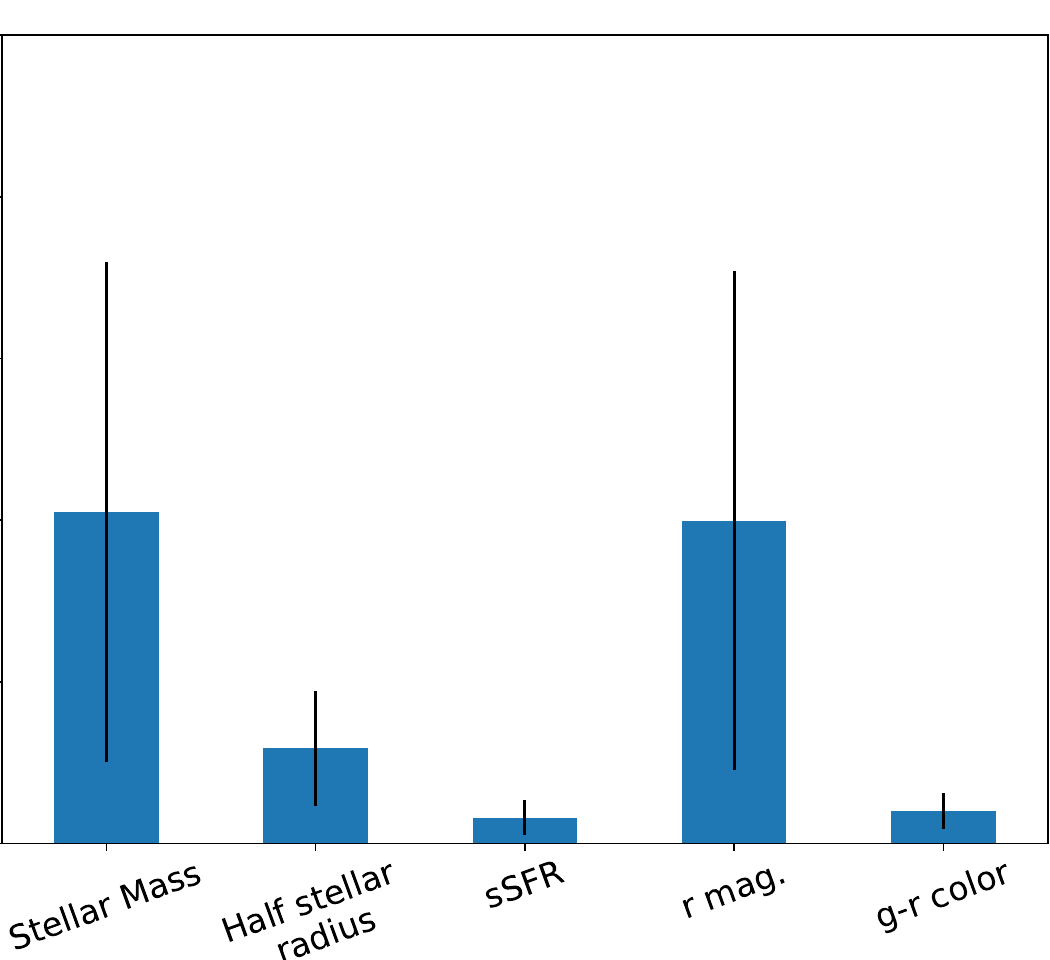}
         \\ \footnotesize Features that are observable in real space.
         \label{fig:group_rf_obs}
      \end{minipage}
        \caption{\footnotesize Random forest feature importance of different combination of central galaxy features in group sized host halos.}
        \label{fig:group_rf}
\end{figure*}

\begin{figure*}[htp!]
     \centering
      \begin{minipage}[t]{0.34\textwidth}
         \centering    \includegraphics[width=\textwidth]{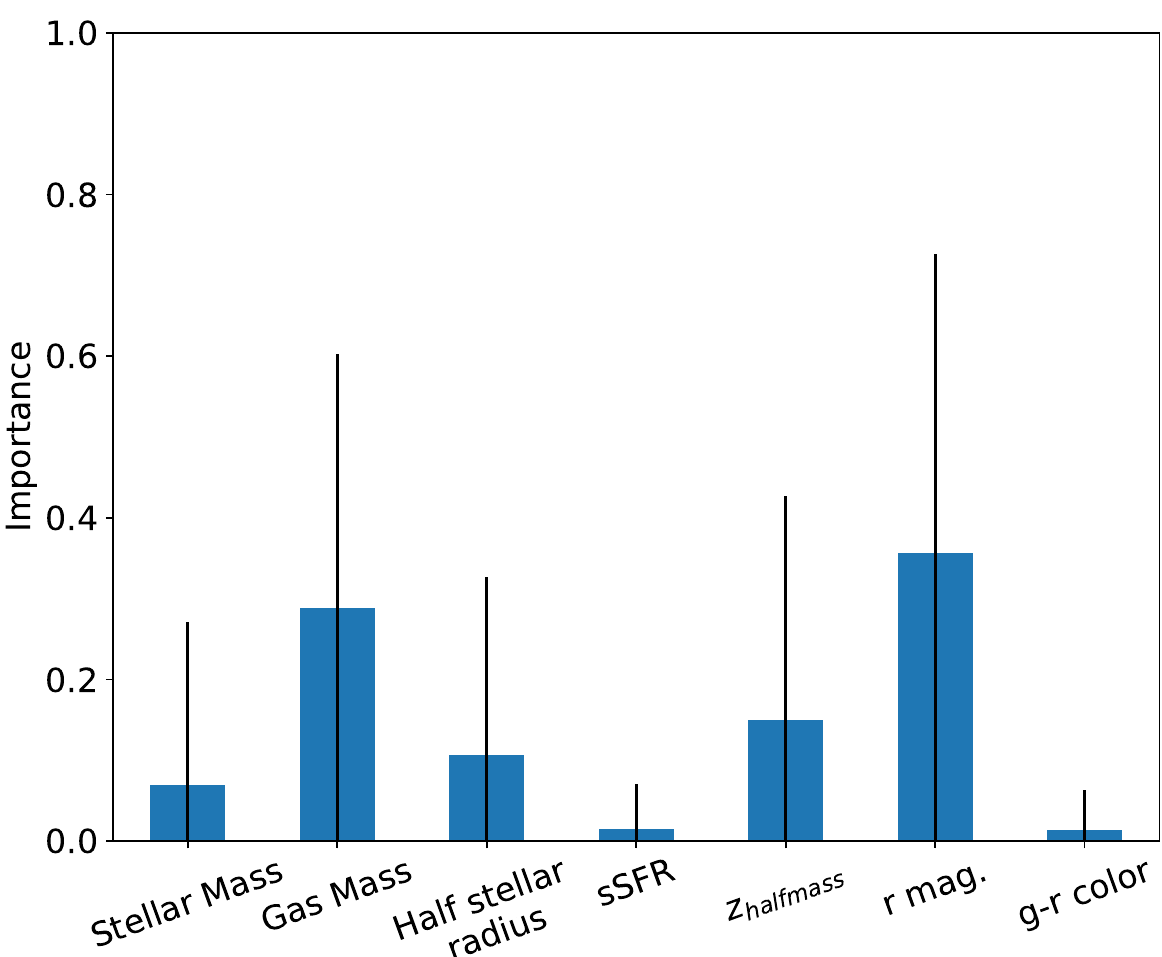}
         \\ \footnotesize All features that could be important for cluster central galaxies.
         \label{fig:cluster_rf_all}
      \end{minipage}
      \begin{minipage}[t]{0.31\textwidth}
         \centering      \includegraphics[width=\textwidth]{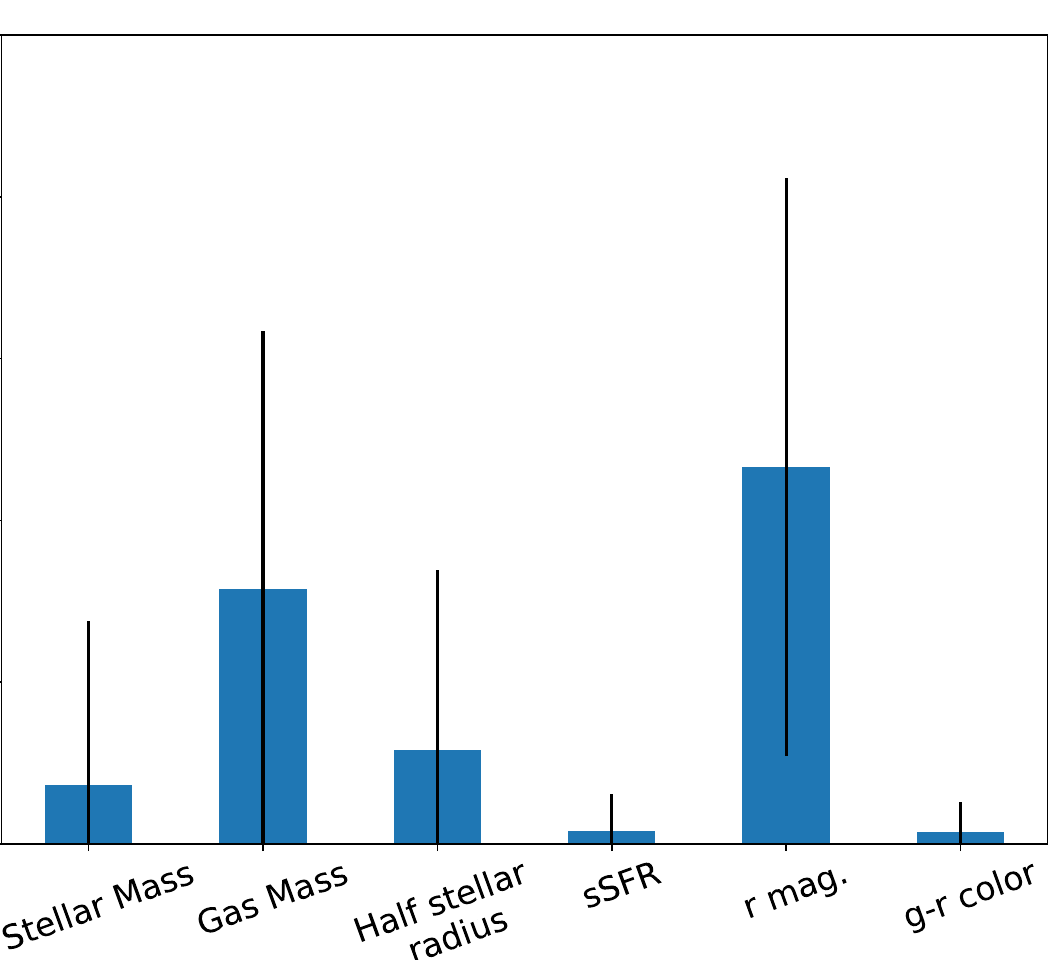}
         \\ \footnotesize Features that are found in simulations.
         \label{fig:cluster_rf_sim}
     \end{minipage} 
      \begin{minipage}[t]{0.31\textwidth}
         \centering    \includegraphics[width=\textwidth]{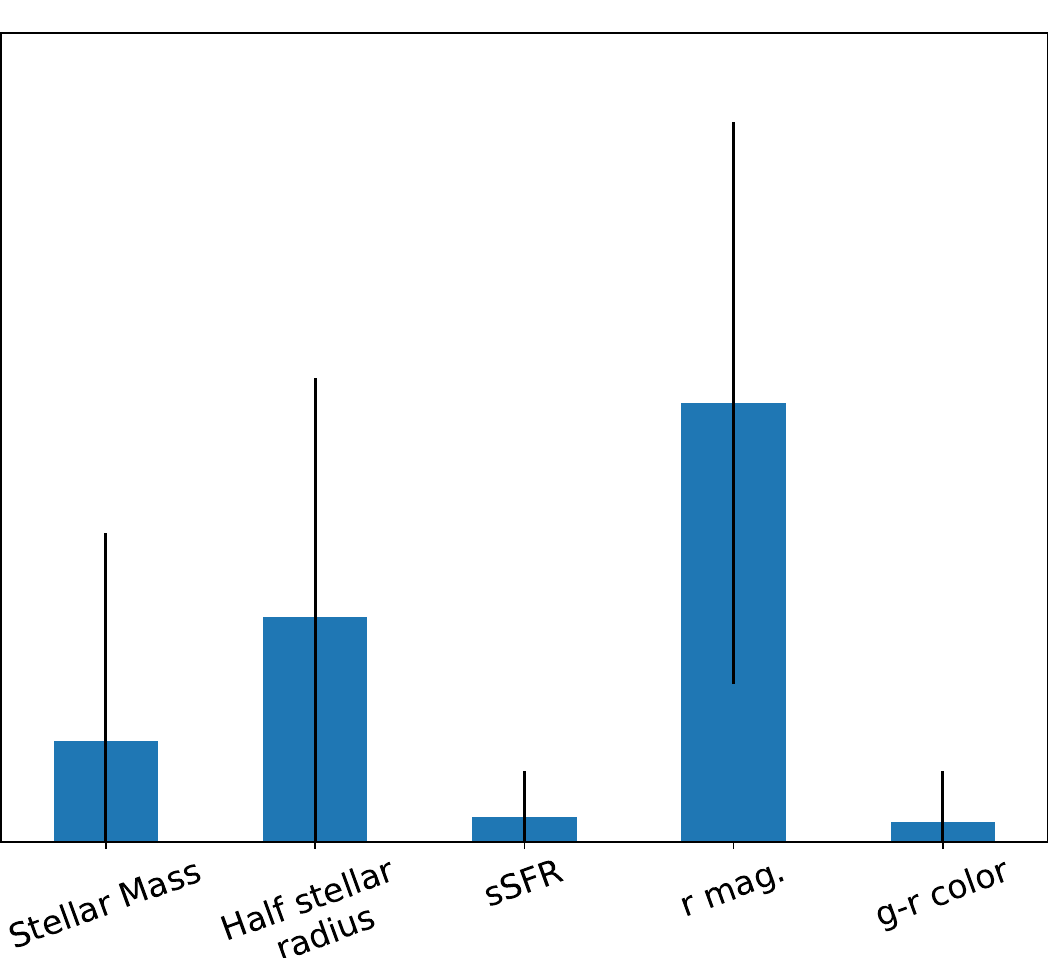}
         \\ \footnotesize Features that are observable in real space.
         \label{fig:cluster_rf_obs}
      \end{minipage}
        \caption{\footnotesize Random forest feature importance of different combination of central galaxy features in cluster sized host halos.}
        \label{fig:cluster_rf}
\end{figure*}

Here, we show the results of some extended error analysis for the host halo mass predictions of central galaxies.  Figures \ref{fig:group_rf} and \ref{fig:cluster_rf} show how features of halos separated into groups and clusters fare in individual categories.  We see that the cluster feature importances with respect to host halo mass differ compared to the groups; gas mass is not nearly as important in clusters compared to groups, while formation redshift importance increases.  Due to the limited number of clusters (17) though, it is unclear whether this result is consistent across clusters in general, or a result of simulation bias.

\section{Error Analysis}

\begin{figure}[ht!]
    \centering
    \includegraphics[width=0.8\linewidth]{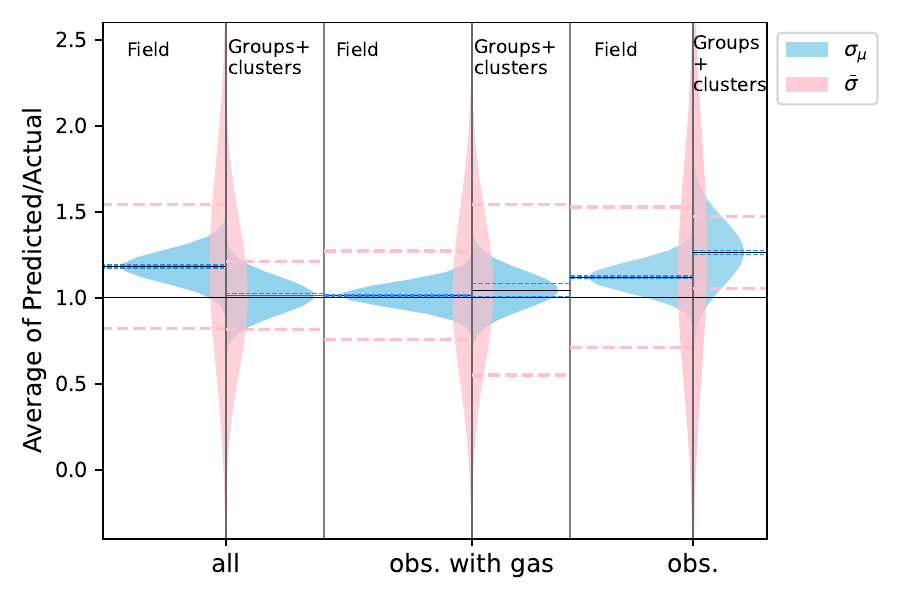}
    \caption{\footnotesize Average and standard deviations of the predicted halo mass errors.  The distributions are centered at the mean, while $\sigma_{\mu}$ is the average standard deviations of random sampling and $\bar{\sigma}$ is the standard deviation of the averages from bootstrapping.  The left panels show the error statistics for field halos, while the right panel is for groups+clusters.}
    \label{fig:violin}
\end{figure}

Figure \ref{fig:violin} and Table \ref{tab:sigma} show the error analysis after bootstrapping over many random sample of halos.  The standard deviations indicated around $20-30\%$ error, which is not ideal, but still within useful limits.  Overall, the mode of the ratio of 
\beq 
f_{\mathrm{error}} = \frac{\mathrm{predicted}\ M_{\mathrm{host}}}{\mathrm{actual}\ M_{\mathrm{host}}}
\label{eqn:ratio}
\eeq
occurs at 1.0, which is expected for accurate predictions.  The standard deviation of Eq. \ref{eqn:ratio}, bootstrapped over randomly selected halos in each mass sample, is given in Table \ref{tab:sigma} and Figure \ref{fig:violin}.  The values in Figure \ref{fig:violin} were obtained by random sampling the test sample of predicted halo masses over 50 iterations, and calculating the mean and standard deviations of these predictions.

\begin{table}[ht!]
    \centering
    \begin{tabular}{|p{0.2\textwidth}|p{0.2\textwidth}|p{0.05\textwidth}|p{0.05\textwidth}|p{0.05\textwidth}|}
    \hline  
    Category & Training Set & $\mu$ & $\bar{\sigma}$ & $\sigma_{\mu}$ \\
    \hline 
    Field & All parameters & 1.2 & 0.36 & 0.0087 \\
    \hline 
    Field & Observables with gas & 1.0 & 0.26 & 0.0055 \\
    \hline  
    Field & Observables & 1.1 & 0.41 & 0.0084 \\
    \hline 
    Group + cluster & All parameters & 1.0 & 0.20 & 0.014 \\
    \hline  
    Group + cluster & Observables with gas & 1.0 & 0.49 & 0.029  \\
    \hline 
    Group + cluster & Observables & 1.3 & 0.21 & 0.015  \\
    \hline 
    \end{tabular}
    \caption{The averages ($\mu$), the average standard deviations of bootstrapping $\sigma_{\mu}$, and standard deviation of the averages from bootstrapping $\bar{\sigma}$. This is the same information as Figure \ref{fig:violin}, but with the numerical values explicitly given.}
    \label{tab:sigma}
\end{table}

\begin{figure*}[ht]
     \centering
      \begin{minipage}[t]{0.32\textwidth}
         \centering    \includegraphics[width=\textwidth]{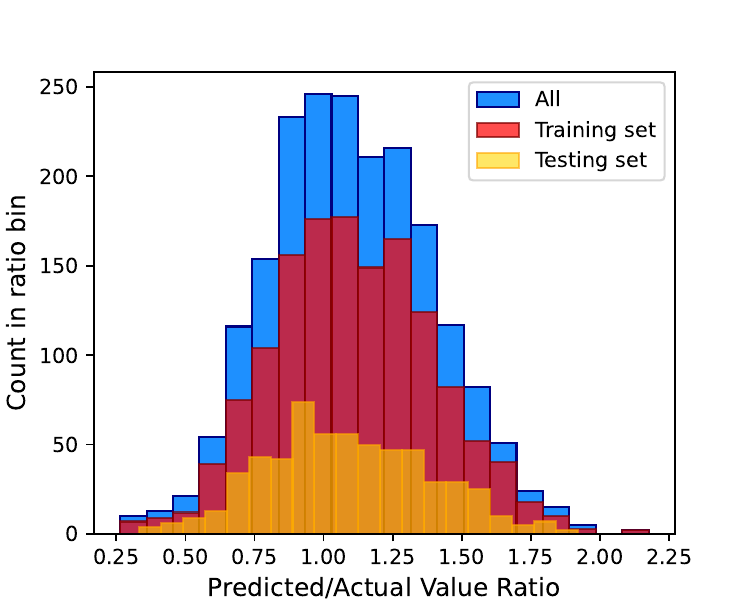}
         \\ \footnotesize All training parameters for central galaxies of field mass range.
         \label{fig:fields_hist_all}
     \end{minipage} 
     \hfill
      \begin{minipage}[t]{0.32\textwidth}
         \centering      \includegraphics[width=\textwidth]{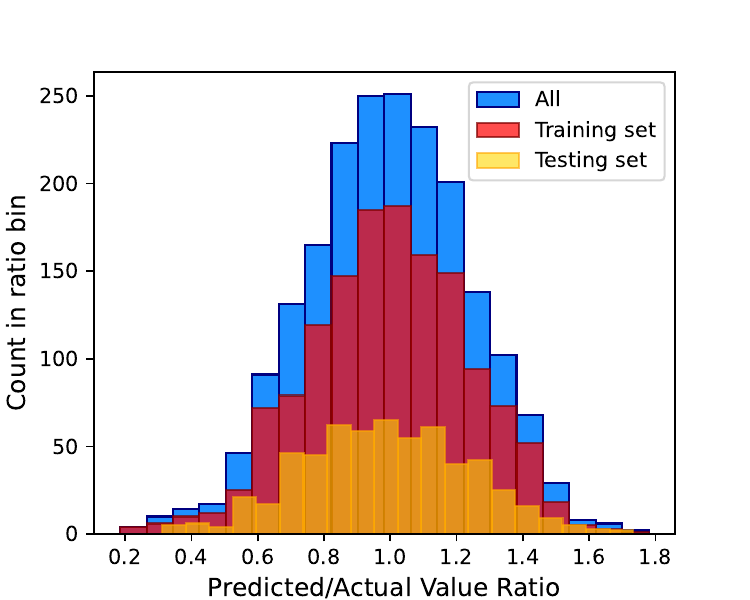}
         \\ \footnotesize Features that are found in simulations.
         \label{fig:fields_hist_sim}
      \end{minipage}
     \hfill
      \begin{minipage}[t]{0.32\textwidth}
         \centering    \includegraphics[width=\textwidth]{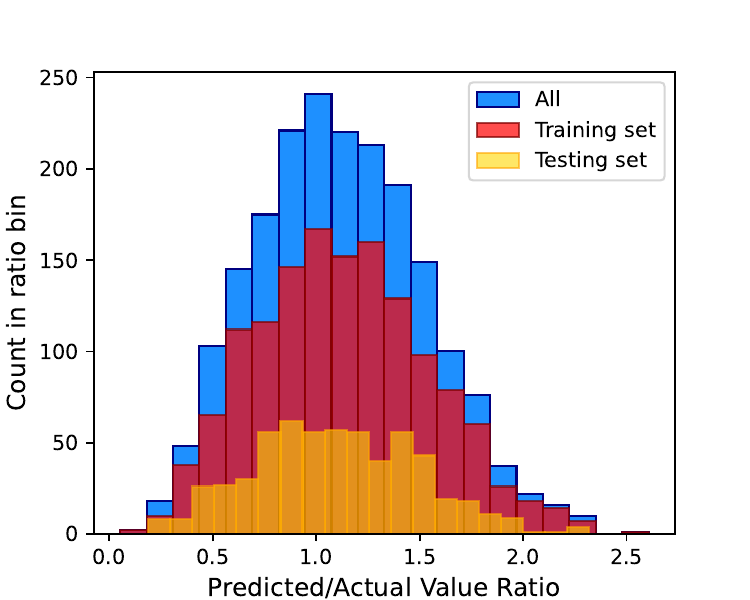}
         \\ \footnotesize Features that are observable in real space.
         \label{fig:fields_hist_obs}
      \end{minipage}
        \caption{\footnotesize The ratio of the predicted host halo masses over the actual halo masses for field mass halos.  For a good prediction, the maximum should be around 1.0 with little deviation, which is what we see here.}
    \label{fig:fields_hist}
\end{figure*}

\begin{figure*}[ht]
     \centering
      \begin{minipage}[t]{0.32\textwidth}
         \centering    \includegraphics[width=\textwidth]{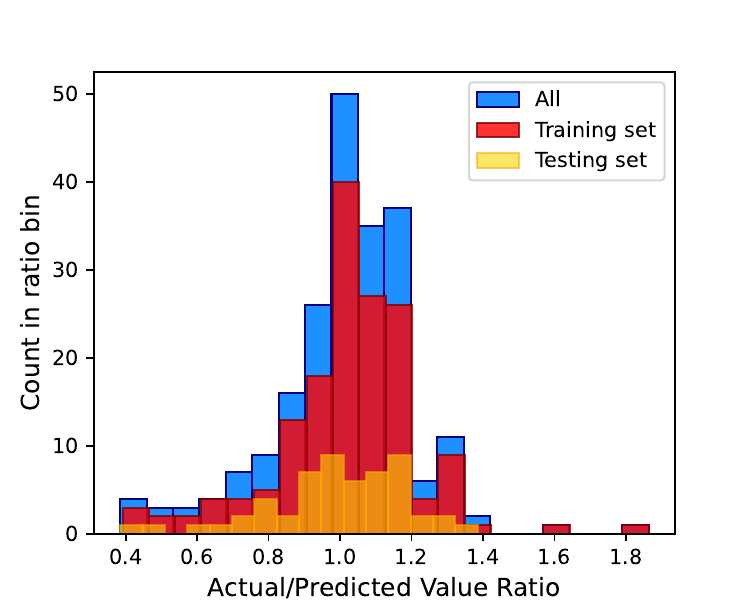}
         \\ \footnotesize All training parameters for central galaxies of group+cluster mass range.
         \label{fig:gc_hist_all}
     \end{minipage} 
     \hfill
      \begin{minipage}[t]{0.32\textwidth}
         \centering      \includegraphics[width=\textwidth]{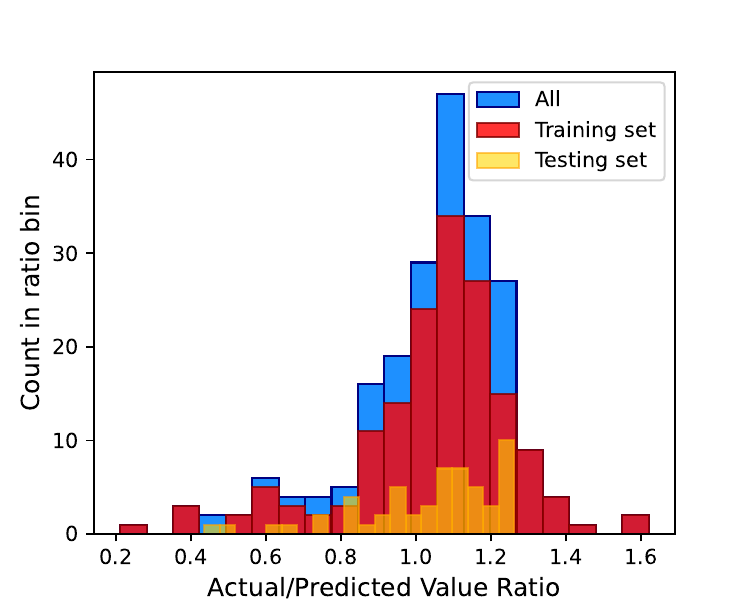}
         \\ \footnotesize Features that are found in simulations.
         \label{fig:gc_hist_sim}
     \end{minipage} 
     \hfill
      \begin{minipage}[t]{0.32\textwidth}
         \centering    \includegraphics[width=\textwidth]{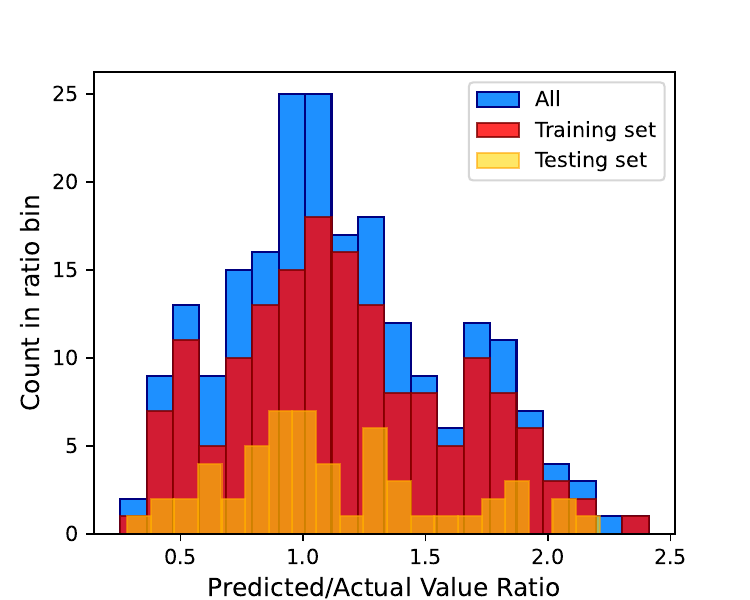}
         \\ \footnotesize Features that are observable in real space.
         \label{fig:gc_hist_obs}
      \end{minipage}
        \caption{\footnotesize The ratio of the predicted host halo masses over the actual halo masses for group + cluster mass halos.  For a good prediction, the maximum should be around 1.0 with little deviation, which is what we see here.}
    \label{fig:gc_hist}
\end{figure*}

We further show standard deviations of training parameters subsets across different prediction algorithms - stellar-halo mass only, OLS, and PySR for pure observables.  Table \ref{tab:goodness_of_fit_host_2} and Figures \ref{fig:fields_hist} and \ref{fig:gc_hist} show that including more parameters consistently outperforms stellar-to-halo mass relation only, while nonlinear scaling relations (PySR) can improve results by a few percent over OLS.  Overall, the results are consistent with what we found in Tables \ref{tab:goodness_of_fit} and \ref{tab:sigma}, and Figure \ref{fig:violin} - there is a high degree of scatter and more parameters improve accuracy (full observables set vs. color, magnitude, and half-stellar radius only).

\begin{table}[ht!]
    \centering
    \begin{tabular}{|c|c|c|c|}
    \hline  
    Category & Galaxy Properties & Training Algorithm & $R^2$ \\
    \hline 
    Field & Observables only & Random Forest & 0.69  \\
    \hline 
    Field & Pure observables & Random Forest & 0.52 \\
    \hline
    Field & Observables only & Least Squares & 0.65 \\
    \hline 
    Field & Pure observables & Least Squares & 0.49 \\
    \hline 
    Field & Stellar Mass only & Least Squares & 0.64 \\
    \hline 
    Field & Observables only & PySR & 0.67 \\
    \hline 
    Field & Pure observables & PySR & 0.53 \\
    \hline 
    Group + cluster & Observables only & Random Forest & 0.77 \\
    \hline 
    Group + cluster & Pure observables & Random Forest & 0.78 \\
    \hline
    Group + cluster & Observables only & Least Squares & 0.83 \\
    \hline 
    Group + cluster & Pure observables & Least Squares & 0.78 \\
    \hline 
    Group + cluster & Stellar Mass only & Least Squares & 0.81 \\
    \hline 
    Group + cluster & Observables only & PySR & 0.85 \\
    \hline 
    Group + cluster & Pure observables & PySR & 0.79 \\
    \hline
    \end{tabular}
    \caption{The $R^2$ values of each sub-category of galaxy observables.  For the pure observables, we've further isolated a subset of just the color, magnitude, and half-light radius.  It can be seen that the pure observable subset does not do as well as the full set of observables we used - we wanted to check that more observables indeed improved results. }
    \label{tab:goodness_of_fit_host_2}
\end{table}

\section{Intrinsic Scatter of Galaxy Properties Across Observations and Different Simulations}
\label{app:scatter}
\begin{figure*}
    \centering
    \includegraphics[trim=0 0 0 15,clip, width=0.9\linewidth]{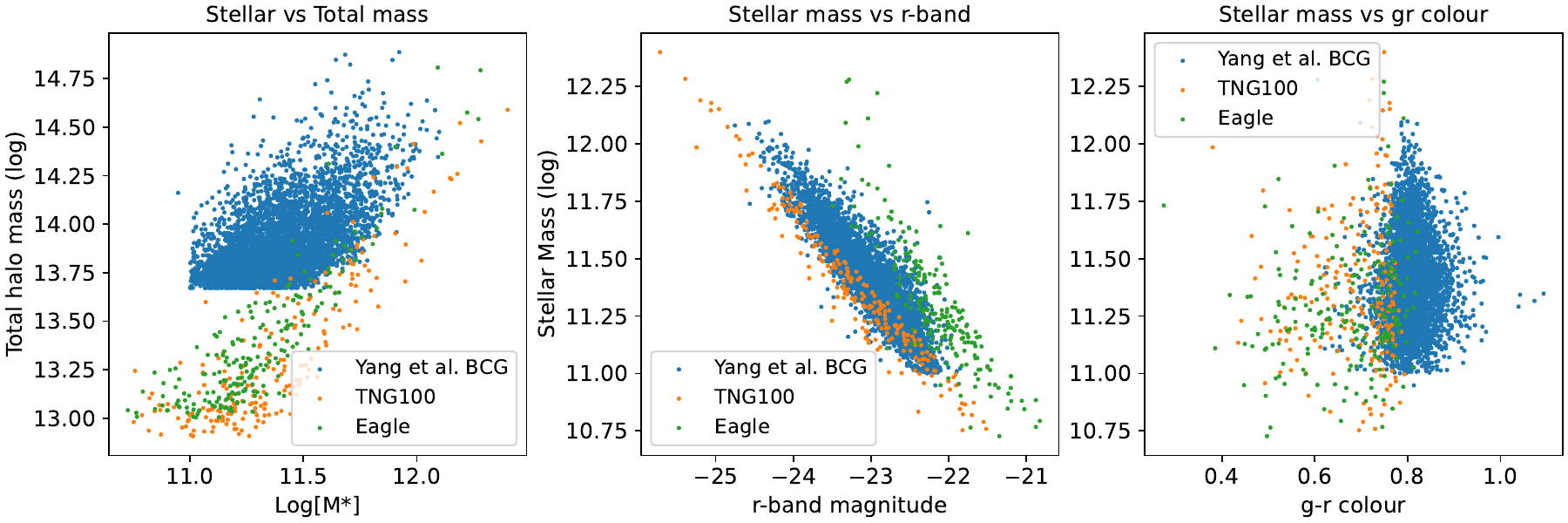}
    \caption{\small Comparisons between stellar mass, halo mass, and observables from \citet{2021ApJ...909..143Y} and simulations (TNG100 and EAGLE).
    \textit{Left Panel:} the distributions of stellar mass vs. total halo mass.
    \textit{Middle Panel:} the distributions of stellar mass vs. r-band magnitude.
    \textit{Right Panel:} the distributions of stellar mass vs $g-r$ color.
    Across all panels, the data points are properties from central galaxies of group and cluster sized halos and the total mass is the halo mass.}
    \label{fig:obs_sims}
\end{figure*}

We see a discrepancy between the estimated observed halo mass and our predictions based on simulation analysis in Sec. \ref{sec:results}, but this could be due to underlying differences in the parameters rather than purely model errors.  We find that observational stellar mass vs. color distributions deviate significantly from both EAGLE and TNG100.  In general, EAGLE had a larger discrepancy than TNG (see Figure \ref{fig:obs_sims}), which is why we stuck with the TNG data for our analysis. This indicates that the discrepancies in color and observed estimates of halo mass distributions persist when going from simulations to observations; the underlying astrophysical causes lie beyond the scope of this work. 

\clearpage
\bibliographystyle{aasjournal}
\bibliography{references}

\end{document}